\documentclass[pdflatex,sn-mathphys-num]{sn-jnl}
\usepackage[T1]{fontenc}
\usepackage{amsmath, amssymb, amsfonts, amsthm}
\theoremstyle{thmstyleone}
\newtheorem{lemma}{Lemma}
\usepackage{graphicx}
\usepackage{xcolor}
\usepackage{booktabs}
\usepackage{placeins}
\usepackage[title]{appendix}
\hypersetup{hidelinks}
\newcommand{\resBARRIERLAMBDA}{4}

\newcommand{\resMINIMAA}{1}
\newcommand{\resNSITES}{8}
\newcommand{\resPHYSICALMASS}{1}
\newcommand{\resSZero}{3.7712}

\newcommand{\resBoxRelDev}{$9.9\times10^{-13}$}
\newcommand{\resBudgetDominant}{$5.1\times10^{-4}$}
\newcommand{\resCTwoCThree}{-12.96}
\newcommand{\resCoarseBarrierHeight}{0.96}
\newcommand{\resCoarseGapRatio}{3.5}

\newcommand{\resCoarseGapRatioResolvedMin}{6.9}
\newcommand{\resCoarseVBarrier}{2.67}
\newcommand{\resCoarseVMin}{1.71}
\newcommand{\resCoarseXBarrier}{0.43}
\newcommand{\resCoarseXMin}{1.29}
\newcommand{\resCritBondExpMax}{-2.23}
\newcommand{\resCritBondExpMin}{-2.51}
\newcommand{\resCritBondTailMax}{-2.12}
\newcommand{\resCritBondTailMin}{-2.20}

\newcommand{\resDEContinuumRef}{0.462428}

\newcommand{\resDEInf}{0.4625}

\newcommand{\resDEInfUncertaintyRel}{$5.4\times10^{-5}$}

\newcommand{\resDESc}{0.2018}
\newcommand{\resDEScOneLoop}{0.6991}
\newcommand{\resDEWkbCorr}{0.6974}

\newcommand{\resDeltaE}{0.3069}
\newcommand{\resDetFitSZeroMin}{6.0}
\newcommand{\resDetOneLoopPrefactor}{3.4641}
\newcommand{\resDetOneLoopRatio}{0.0833}
\newcommand{\resDetRatioExactOverLeading}{3.4623}

\newcommand{\resEstFftBiasSixPeriodsPct}{0.42}
\newcommand{\resEstFftBiasTierEPct}{6.1}
\newcommand{\resEstFitErrMax}{$3.5\times10^{-9}$}
\newcommand{\resEstMethodTwoFftErrPct}{4.1}
\newcommand{\resEstMethodTwoFitErr}{$5.0\times10^{-10}$}
\newcommand{\resEstMethodTwoPeriods}{1.95}
\newcommand{\resFidTwoTwob}{1.0000}
\newcommand{\resInitGaussWeightMin}{0.91}
\newcommand{\resInitHwFitErrLPct}{0.34}
\newcommand{\resInitHwFitErrGaussPct}{0.9}
\newcommand{\resInitHwFitErrSitePct}{8.7}
\newcommand{\resInitSiteErrLongPct}{1.7}
\newcommand{\resInitSiteWeightNEight}{0.74}
\newcommand{\resInitSiteWeightLast}{0.05}

\newcommand{\resIsingValidationErr}{$4.1\times10^{-6}$}
\newcommand{\resLooSpreadRel}{$4.4\times10^{-5}$}

\newcommand{\resMagicPeriodLogSlope}{2.02}

\newcommand{\resOmegaSc}{5.6569}

\newcommand{\resRankMaxIntermediate}{4}
\newcommand{\resRankMaxStepEnd}{2}

\newcommand{\resRichardsonVsRefRel}{$5.4\times10^{-5}$}
\newcommand{\resSolverRel}{$3.8\times10^{-13}$}
\newcommand{\resSplitBondMpsMaxDevSteps}{0.8}
\newcommand{\resSplitEoCostRatioOneTwoEight}{1.8}
\newcommand{\resSplitEoCostRatioSixFour}{0.92}
\newcommand{\resSplitEoCrossoverN}{70}
\newcommand{\resSplitEoDtRatioMax}{1.5}
\newcommand{\resSplitEoDtRatioMin}{1.2}
\newcommand{\resSplitEoExponent}{-2.11}
\newcommand{\resSplitOrderFourNetMax}{1.7}
\newcommand{\resSplitOrderFourNetMin}{1.0}
\newcommand{\resSplitOrderFourRelaxMax}{9}
\newcommand{\resSplitOrderFourRelaxMin}{5}
\newcommand{\resSplitSweepCrossoverN}{26}
\newcommand{\resSplitTvDtMax}{0.085}
\newcommand{\resSplitTvDtMin}{0.065}
\newcommand{\resSplitTvExponent}{-0.04}

\newcommand{\resTierECostExponentRefined}{3.3}

\newcommand{\resTierEDtStarExponentRefined}{-2.28}
\newcommand{\resTierEDtStarExponentRefinedErr}{0.09}
\newcommand{\resTierEDtStarExponentTail}{-2.12}
\newcommand{\resTierEDtStarExponentTailErr}{0.03}

\newcommand{\resTierEWorstRel}{$6.6\times10^{-5}$}

\newcommand{\resRatioInfSc}{2.29}
\newcommand{\resRatioExactSc}{1.52}
\newcommand{\resTtunnel}{10.24}

\newcommand{\resRatioInfOneLoop}{0.6615}
\newcommand{\resOneLoopOverExactPct}{51}
\newcommand{\resTierEContinuumRel}{$6.3\times10^{-5}$}
\newcommand{\resDEInfQuoted}{0.46245(3)}

\newcommand{\resRichardsonNmax}{192}
\newcommand{\resRichardsonWindow}{48\ldots192}
\newcommand{\resBoxLconverged}{2.5}
\newcommand{\resRichardsonWindowLowRel}{1.5\%}
\newcommand{\resRichardsonReferenceDx}{0.005}

\newcommand{\resCsNFirst}{12}
\newcommand{\resCsNLast}{50}
\newcommand{\resCsBudgetFirst}{15}
\newcommand{\resCsNeededFirst}{137}
\newcommand{\resCsBudgetLast}{3}
\newcommand{\resCsNeededLast}{1918}
\newcommand{\resCsFixedStepFactorLast}{14}

\newcommand{\resConvRelErrNSites}{34}

\newcommand{\resConvRelErrFirstFit}{1.2}

\newcommand{\resXhi}{3}
\newcommand{\resXdomain}{[-3,3]}

\newcommand{\hwBackend}{ibm\_kingston}

\newcommand{\hwNSites}{8}
\newcommand{\hwDt}{0.55}
\newcommand{\hwShots}{4096}
\newcommand{\hwPoints}{14}
\newcommand{\hwTwoQError}{1.86}
\newcommand{\hwDeltaE}{0.30487}
\newcommand{\hwDeltaEErr}{0.00127}
\newcommand{\hwDevPct}{0.67}
\newcommand{\hwTrotterPct}{0.34}
\newcommand{\hwDevicePct}{1.00}
\newcommand{\hwAmplitude}{0.863}
\newcommand{\hwTmax}{12.7}
\newcommand{\hwTwoQPerStep}{14}
\newcommand{\hwTwoQPrep}{32}
\newcommand{\hwTwoQMax}{354}

\newcommand{\hwScanTol}{0.02}
\newcommand{\hwScanTwelveDtMin}{0.20}
\newcommand{\hwScanTwelveTwoQ}{924}
\newcommand{\hwIdealMaxDev}{0.012}

\newcommand{\hwDeltaERaw}{0.31105}
\newcommand{\hwDevPctRaw}{1.34}
\newcommand{\hwAmpRaw}{0.678}
\newcommand{\hwLeakTau}{24.2}
\newcommand{\hwLeakTauErr}{0.8}
\newcommand{\hwLeakIntercept}{0.803}
\newcommand{\hwLeakLast}{0.481}
\newcommand{\hwBackendTwo}{ibm\_fez}

\newcommand{\hwTwoQErrorTwo}{2.83}

\newcommand{\hwAmpTwo}{0.770}
\newcommand{\hwAmpTwoSd}{0.033}
\newcommand{\hwRepNTwo}{5}
\newcommand{\hwJointDeltaEPct}{7.0}

\newcommand{\hwLamMin}{2}
\newcommand{\hwLamMax}{6}
\newcommand{\hwSlope}{0.803}
\newcommand{\hwSlopeErr}{0.008}
\newcommand{\hwSlopeExact}{0.829}
\newcommand{\hwSlopeExactW}{0.812}
\newcommand{\hwSlopeExactErr}{0.056}
\newcommand{\hwSlopeTheory}{1.886}
\newcommand{\hwZneDeltaE}{0.29918}
\newcommand{\hwZneDevPct}{2.53}
\newcommand{\hwZneAmp}{1.014}

\newcommand{\hwValCohTwoQ}{354}

\newcommand{\hwDevDeltaESigma}{3.3}
\newcommand{\hwDevDeltaEPct}{3.17}
\newcommand{\hwDevAmpKing}{0.853}
\newcommand{\hwDevAmpKingSd}{0.012}
\newcommand{\hwDevAmpSigma}{5.5}

\newcommand{\hwDevDeltaESigmaSess}{2.1}
\newcommand{\hwDevAmpSigmaSess}{6.0}
\newcommand{\hwDevDeltaEPSess}{0.10}
\newcommand{\hwDevAmpPSess}{0.02}
\newcommand{\hwDevDeltaEDofSess}{4.0}
\newcommand{\hwDevAmpDofSess}{2.3}
\newcommand{\hwSessN}{4}
\newcommand{\hwSessNTwo}{3}
\newcommand{\hwSessWithinPct}{0.27}
\newcommand{\hwSessBetweenPct}{2.6}
\newcommand{\hwRepMeanSePct}{1.3}
\newcommand{\hwRepMeanSe}{0.00397}
\newcommand{\hwLayMatchedN}{9}
\newcommand{\hwLayMatchedShiftPct}{0.03}
\newcommand{\hwZneSigma}{1.1}
\newcommand{\hwZneAmpSigma}{13}
\newcommand{\hwSamplerDevMean}{+0.26}
\newcommand{\hwSamplerDevSd}{2.60}
\newcommand{\hwIsNzeroPostMean}{846}
\newcommand{\hwIsNzeroPostSd}{127}
\newcommand{\hwIsNzeroPostRelPct}{15}
\newcommand{\hwIsNbelow}{1}
\newcommand{\hwIsRatioMean}{4.15}
\newcommand{\hwIsRatioRound}{4}
\newcommand{\hwIsNzeroRawMean}{204}
\newcommand{\hwIsNzeroRawSd}{18}
\newcommand{\hwIsRatioSd}{0.46}
\newcommand{\hwIsRatioRelPct}{11}
\newcommand{\hwSlopeDriftErr}{0.029}

\newcommand{\hwValIprSpreadPct}{6}
\newcommand{\hwValIprRise}{4}
\newcommand{\hwVerifiedSpan}{104}
\newcommand{\hwSecTvdHigh}{0.06}
\newcommand{\hwSecTvdMid}{0.12}
\newcommand{\hwSecTvdLow}{0.21}
\newcommand{\hwSecNPoints}{196}
\newcommand{\hwSecNLow}{9}
\newcommand{\hwKingPsubMin}{0.43}
\newcommand{\hwLayChainsRef}{3}
\newcommand{\hwLayMismatchRef}{5}
\newcommand{\hwLayMismatchRep}{1}
\newcommand{\hwDriftSeenH}{7.4}
\newcommand{\hwValGapMin}{133}
\newcommand{\hwRawAbsDevMean}{2.2}
\newcommand{\hwPostAbsDevMean}{2.4}
\newcommand{\hwExclRiseSigma}{19}
\newcommand{\hwRiseMaxOther}{5}
\newcommand{\hwRepRelPctExcl}{2.86}
\newcommand{\hwRepN}{9}
\newcommand{\hwRepMean}{0.30692}
\newcommand{\hwRepSd}{0.00823}
\newcommand{\hwRepRelPct}{2.68}
\newcommand{\hwRepRangePct}{6.39}
\newcommand{\hwRepStatPct}{0.40}

\newcommand{\hwRepInflation}{6.8}
\newcommand{\hwSessRelPct}{0.35}
\newcommand{\hwSessStatPct}{0.39}
\newcommand{\hwSessChi}{0.7}
\newcommand{\hwDriftPct}{3.24}
\newcommand{\hwSessTwoChi}{0.2}
\newcommand{\hwSessShiftPct}{6.1}

\newcommand{\hwAblDevPct}{-0.38}
\newcommand{\hwAblDevPctOn}{+2.16}
\newcommand{\hwAblAmp}{0.854}
\newcommand{\hwAblAmpOn}{0.830}
\newcommand{\hwAblNzero}{821}
\newcommand{\hwAblNzeroOn}{608}
\newcommand{\hwLeakKingMean}{2.02}
\newcommand{\hwLeakKingSd}{0.29}

\newcommand{\hwIsBackend}{ibm\_kingston}

\newcommand{\hwIsN}{8}
\newcommand{\hwIsDt}{0.40}
\newcommand{\hwIsJ}{1}
\newcommand{\hwIsG}{0.5}
\newcommand{\hwIsH}{0.25}
\newcommand{\hwIsShots}{4096}
\newcommand{\hwIsTwoQMax}{322}
\newcommand{\hwIsTwoQPerStep}{14}
\newcommand{\hwIsTwoQOffset}{32}
\newcommand{\hwIsTmax}{9.2}
\newcommand{\hwIsMzFv}{-0.105}
\newcommand{\hwIsMzCtl}{-0.490}
\newcommand{\hwIsMzFvIdeal}{-0.203}
\newcommand{\hwIsMzCtlIdeal}{-0.766}
\newcommand{\hwIsSepSigma}{33}
\newcommand{\hwIsNzeroFv}{455}
\newcommand{\hwIsNzeroFvErr}{10}
\newcommand{\hwIsNzeroCtl}{651}
\newcommand{\hwIsNzeroCtlErr}{9}
\newcommand{\hwIsResidUnRaw}{0.154}
\newcommand{\hwIsResidCtl}{0.037}
\newcommand{\hwIsIprFv}{46}
\newcommand{\hwIsIprCtl}{8.5}
\newcommand{\hwIsIprRatio}{5.4}
\newcommand{\hwIsDwFv}{0.284}
\newcommand{\hwIsDwFvIdeal}{0.157}
\newcommand{\hwIsDwCtl}{0.200}
\newcommand{\hwIsDwCtlIdeal}{0.037}
\newcommand{\hwIsRmsRaw}{0.093}
\newcommand{\hwIsRmsCorr}{0.034}
\newcommand{\hwIsRmsFactor}{2.8}
\newcommand{\hwFixN}{4}
\newcommand{\hwFixPairDiffPct}{0.25}
\newcommand{\hwFixPairGapMin}{2}
\newcommand{\hwFixMinPct}{4.0}
\newcommand{\hwFixMaxPct}{8.9}
\newcommand{\hwFixHourShiftPct}{2.4}
\newcommand{\hwFixHourGapMin}{64}
\newcommand{\hwFreeVsFixPct}{3.0}
\newcommand{\hwFreeGapMin}{3}
\newcommand{\hwGaussDiffMin}{-1.6}
\newcommand{\hwGaussDiffMax}{+0.8}
\newcommand{\hwGaussDiffAbsMax}{1.6}
\newcommand{\hwGaussIdealDiff}{+0.6}

\begin{document}

\title[Instanton tunnelling as a verifiable benchmark]{Instanton tunnelling as a verifiable benchmark for digital quantum simulation: circuit structure, classical simulability and hardware reproducibility}

\author*[1]{\fnm{Gerhard} \sur{Hellstern}}\email{gerhard.hellstern@dhbw-stuttgart.de}
\author[2]{\fnm{Marc} \sur{Maussner}}

\affil*[1]{\orgname{Duale Hochschule Baden-W\"urttemberg (DHBW) Stuttgart}, \orgaddress{\city{Stuttgart}, \country{Germany}}}
\affil[2]{\orgname{infoteam Software AG}, \orgaddress{\city{Bubenreuth}, \country{Germany}}}

\abstract{We use instanton-mediated tunnelling in the quartic double well as a
benchmark for digital quantum simulation that is checkable against an exact
answer at every step. The instanton enters through the classical analysis; the
hardware tests the dynamics of the lattice model. A unary (single-excitation) encoding maps $N$ lattice sites to $N$
qubits, with the single-excitation dynamics as a reference that remains classically
cheap at any $N$. For a second-order Trotter circuit with
adjacent, equal-angle $XX$/$YY$ pairs on each bond, we prove that the Schmidt
rank across any cut is at most four at every gate boundary and at most two
after each pair, independently of $N$. Matrix-product-state simulations of this
gate list reproduce the exact splitting up to $N=48$. For this bond-wise formula the admissible Trotter step shrinks roughly as
$N^{-2}$ ($N^{\resTierEDtStarExponentRefined}$ over $N=16$--$64$ for our accuracy
criterion, $N^{\resCritBondTailMin}$ to $N^{\resCritBondTailMax}$ for $N\ge24$
under all criteria tested), so one tunnelling
period costs about $N^{\resTierECostExponentRefined}$ two-qubit gates; a
kinetic/potential splitting keeps the step near $\resSplitTvDtMin$ up to
$N=128$, so this scaling belongs to splitting the kinetic term bond by bond. At a
coarser, separately bounded step we measure the full transfer at
$N=\hwNSites$ on two superconducting processors, recovering the lattice
splitting to a few percent and its dependence on the barrier height, which at
this resolution is set by the lattice rather than by the instanton.
Over $\hwRepN$ repetitions, runs taken minutes apart agree to shot noise,
while runs hours to weeks apart differ by up to $\hwSessShiftPct\,\%$, also on a
fixed set of qubits;
with sessions as the sampling unit ($\hwSessN$ and $\hwSessNTwo$ sessions),
the processors differ in the damping ($p\approx\hwDevAmpPSess$), while a
difference in the splitting is not established ($p\approx\hwDevDeltaEPSess$). Normalising by the measured single-excitation
population extends the usable depth about $\hwIsRatioRound$-fold but does not
improve the splitting. A Gaussian start state, which needs no diagonalisation,
reproduces the splitting of the eigenvector-based start to within
$\hwGaussDiffAbsMax$ percentage points.}


\maketitle

\section{Introduction}\label{sec:intro}
Instantons are classical solutions to the equations of motion in Euclidean 
spacetime with finite action, and they provide the bridge between classically 
disconnected vacuum states. In standard quantum mechanics, the phenomenon of
tunnelling across a potential barrier is often described using the WKB approximation; 
however, the instanton formalism, based on Feynman's path integral, offers a 
more robust framework that extends naturally to Quantum Field Theory (QFT) and 
Gauge Theories \cite{garg_2000}.

The transition from a real-time formulation to an imaginary-time (Euclidean) 
formulation transforms the tunnelling problem into a classical propagation 
problem in an inverted potential. This rotation allows for the identification 
of a well-defined ``path'' -- the instanton -- that captures the non-perturbative 
nature of the vacuum. Unlike perturbative effects, which can be expanded in 
powers of $\hbar$, instanton-induced effects are proportional 
to $\exp(-S_0/\hbar)$, introducing an essential singularity at $\hbar = 0$.

Digital quantum simulation suits non-perturbative problems because the device
carries the state itself and its cost is set by circuit depth, not by
entanglement. Classical state-based methods meet specific obstructions instead:
entanglement growth for tensor networks, the sign problem for quantum Monte
Carlo. Neither is present in the model studied here, which is why we treat it as
a benchmark and not as a candidate for advantage. The encoding of the coordinate
$x$ into qubits determines the gate cost and which errors are detectable; we use
a unary (one-hot) encoding, whose single excitation makes leakage out of the
physical subspace directly observable.

The results are obtained at seven levels of computation (Table~\ref{tab:comp_tiers}),
and each result is labelled by its tier. Tiers~A--C and~E address the
double-well Hamiltonian classically, so each of their numbers can be checked
against an exact answer. Tier~D is a projection from gate counts and an error
model. Tier~D$'$ is the device measurement, whose analysis is reproducible from
frozen raw counts but whose runs are not. Tier~F is a control on a different
Hamiltonian. The relative uncertainties of all tiers are collected in
Table~\ref{tab:budget} (Appendix~\ref{app:budget}).

\begin{table}[htbp]
\centering
\begin{tabular}{|c|p{3.0cm}|p{5.0cm}|p{2.3cm}|}
\hline
\textbf{Tier} & \textbf{Method} & \textbf{Description} & \textbf{Sections} \\
\hline
A & Analytical / semiclassical & Zero-mode-only and one-loop instanton formulas, WKB, DIGA & \ref{sec:theory}, \ref{sec:semiclassical_comparison}, App.~\ref{app:semiclassics} \\
B & Classical exact computation & Diagonalisation of the $N\!\times\!N$ unary Hamiltonian on a classical computer & \ref{sec:exact_diag_results}, \ref{sec:semiclassical_comparison} \\
C & Classical simulation of quantum circuits & Qiskit noiseless statevector simulator executing Trotterised circuits & \ref{sec:quantum_circuits} \\
D & Projected quantum hardware & Resource estimates and fidelity projections for NISQ / fault-tolerant devices, and a preparatory study on a calibrated noise model & \ref{sec:circuit_scaling}, Suppl.~S8 \\
D$'$ & Device measurement & Repeated survival-probability measurements on two superconducting processors, analysed from frozen raw counts by the same code & \ref{sec:hw_campaign}, App.~\ref{app:hw_repro}, \ref{app:hw_control} \\
E & Tensor-network circuit simulation & The same Trotterised gate list contracted as a matrix product state (quimb), at sizes where a dense statevector is unavailable & \ref{sec:tier_e} \\
F & Control system & A different Hamiltonian --- an Ising chain undergoing false-vacuum decay --- simulated with the same tools, to show which conclusions are properties of the model rather than of the method & App.~\ref{sec:tier_f}, \ref{app:hw_control} \\
\hline
\end{tabular}
\caption{The seven computational tiers. Tier~B (exact diagonalisation) is the reference for all others.}
\label{tab:comp_tiers}
\end{table}

Quantum simulation of tunnelling has been approached from several directions. 
Abel, Chancellor and Spannowsky~\cite{abel_2021} encode a scalar field theory on a quantum annealer
and recover the tunnelling solution, a static problem rather than real-time dynamics. Trapped-ion experiments have prepared 
superpositions of topological defects directly~\cite{cheng_2024}, and false-vacuum decay in Ising-type 
spin chains has been studied both numerically and on hardware~\cite{lagnese_2021,milsted_2022,kormos_2017},
recently on a quantum annealer~\cite{vodeb_2025} and a cold-atom gauge-theory simulator~\cite{zhu_2026}; 
those are many-body problems in which entanglement grows and the interesting question is cost. 
Jordan, Lee and Preskill~\cite{jordan_2012} give the asymptotic algorithmic framework for scalar field theory 
into which such work eventually has to fit.

Closest to the present model is a line of work that digitises the same single-particle tunnelling problem 
directly. Sornborger~\cite{sornborger_2012} simulates tunnelling in a one-dimensional potential using a 
\emph{binary} encoding, $n = \log_2 N$ qubits for $N$ lattice points, and Feng et al.~\cite{feng_2013} 
realised that algorithm on NMR hardware --- to our knowledge the first experimental digital simulation of
barrier tunnelling. More recently, McGarry et al.~\cite{mcgarry_2026} observe wavepacket tunnelling in 
programmable double-well potentials on a trapped-ion processor by a continuous-variable route, using bosonic
quantum signal processing rather than a Trotter product formula. The binary encoding needs exponentially fewer
qubits; the unary one provides the observable subspace constraint used in Sections~\ref{sec:quantum_circuits}
and~\ref{sec:tier_e} and the bond-dimension bound of Lemma~\ref{prop:chi4}.
Grid-based simulation of single-particle dynamics has a longer history: the split-operator
method alternates the potential with the kinetic term, diagonalised by a quantum Fourier
transform in a binary register~\cite{zalka_1998,kassal_2008,benenti_strini_2008}, and the
complexity of real-space dynamics has been analysed in detail~\cite{childs_realspace_2022}.
Unary and binary encodings of bosonic modes, and the digitisation error of a field on a finite
grid, have been compared in Refs.~\cite{macridin_2018,klco_savage_2019}.
The trade-off between unary, binary and Gray-code encodings of $d$-level systems has been analysed
systematically for gate counts and qubit numbers~\cite{sawaya_2020,di_matteo_2021}, and unary
encodings have been used for their simple, excitation-preserving circuits in other
applications~\cite{ramos_calderer_2021}; we use the encoding for the observable sector it provides.
Checking a conserved quantity to detect and mitigate errors is established as symmetry
verification~\cite{bonet_monroig_2018,mcardle_2019,sagastizabal_2019}; the hardware estimator of
Section~\ref{sec:hw_campaign} is a variant of it in which the sector population is measured by a
separate circuit.

The model is single-particle and classically easy, so nothing here bears on
quantum advantage. What the paper offers is a gate-level pipeline whose classical
tiers are each checkable against an exact answer. The paper makes four contributions.
(i)~\emph{Verification:} the one-loop fluctuation determinant of the quartic double well,
$\det{}'/\det_0 = 1/12$~\cite{escobar_ruiz_2015}, is confirmed along two independent
numerical routes, which agree to \resBudgetDominant; it separates the zero-mode-only
and one-loop predictions from the exact lattice splitting
(Section~\ref{sec:determinant_result}).
(ii)~\emph{Structural:} for the Trotter synthesis used here we show that the Schmidt rank never
exceeds four at any gate boundary and two after each pair of $XX$ and $YY$ rotations,
independently of $N$ (Lemma~\ref{prop:chi4}).
(iii)~\emph{Numerical:} for the bond-wise product formula that Qiskit synthesises, the
admissible Trotter step shrinks as
$N^{\resTierEDtStarExponentRefined\pm\resTierEDtStarExponentRefinedErr}$ over $N=16$--$64$, and
as $N^{\resCritBondTailMin}$ to $N^{\resCritBondTailMax}$ for $N\ge24$ under every accuracy
criterion tested, so a tunnelling period costs about $N^{\resTierECostExponentRefined}$
two-qubit gates in that range,
whereas a kinetic/potential splitting removes the $N$ dependence of the step;
the odd/even bond ordering of the device circuit scales in the same way with a
step larger by a factor \resSplitEoDtRatioMin--\resSplitEoDtRatioMax\
(Section~\ref{sec:tn_dt_scaling}); the bound of (ii) is attained in every case tested.
(iv)~\emph{Experimental:} a reproducibility-qualified measurement of the complete transfer on the coarse $N=\hwNSites$ lattice, whose exact splitting is the reference (Section~\ref{sec:n8_scope}), on two superconducting processors.
Full transfers on small registers have been shown before~\cite{feng_2013,mcgarry_2026}; new is the
repeated, session-resolved analysis, which shows the uncertainty to be dominated by differences
between sessions, not by shot noise (Section~\ref{sec:hw_campaign}). We regard (iv), together with the benchmark
itself, as the main contribution.

\section{Semiclassical Theory of the Double Well}\label{sec:theory}

The standard semiclassical derivations are reproduced in Sections~S1--S6 of the Supplementary Information; Appendix~\ref{app:semiclassics} checks the determinant numerically. This section fixes the potential, the notation and the reference formulas.

\subsection{Double-Well Potential and Instanton Solution}\label{sec:double_well}
We consider the quartic potential:
\begin{equation}
V(x) = \lambda (x^2 - a^2)^2
\end{equation}
with two degenerate minima at $x=\pm a$, coupled by tunnelling.

\begin{figure}[htbp]
\centering
\includegraphics[width=0.75\textwidth]{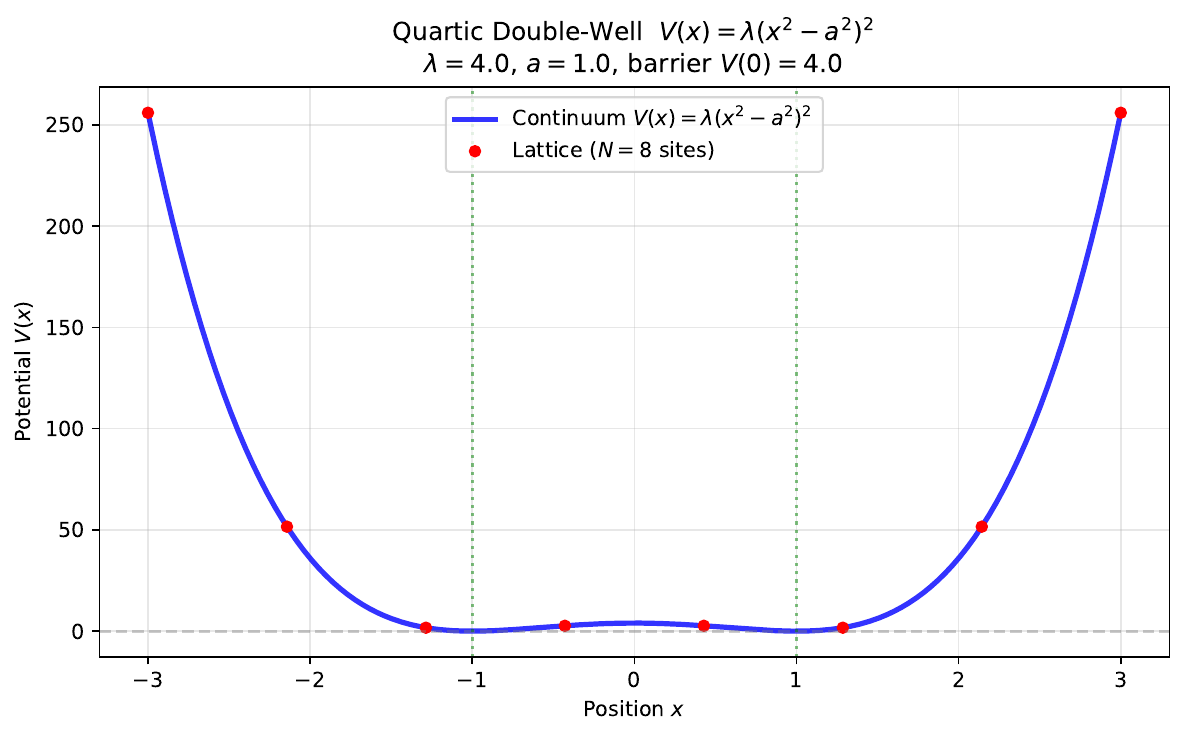}
\caption{The quartic double-well potential $V(x) = \lambda(x^2-a^2)^2$ for $\lambda=4$, $a=1$ (line) and its discretisation on $N=8$ sites (markers). The barrier height is $V(0)=\lambda a^4=4$. At $N=8$ the minima are not grid points and one pair of points spans the barrier (Section~\ref{sec:n8_scope}).}
\label{fig:double_well}
\end{figure}

In the Euclidean formulation, the ``motion'' occurs in the potential $-V(x)$, where $x = \pm a$ are the ``hilltops'' (see Figure~\ref{fig:double_well}). A classical solution that starts at $x = -a$ at $\tau \to -\infty$ and arrives at $x = +a$ at $\tau \to +\infty$ is called an instanton.

The barrier height is $V(0) = \lambda a^4$, and the curvature at each minimum determines the oscillator frequency:
\begin{equation}
\omega = \sqrt{\frac{V''(\pm a)}{m}} = \sqrt{\frac{8\lambda a^2}{m}}
\end{equation}
It sets the time scale of the intra-well oscillations and the width of the instanton $x(\tau) = a\tanh(\omega(\tau-\tau_0)/2)$ (Supplementary Information, Section~S3).

\subsection{Semiclassical Reference Formulas}\label{sec:coleman_formula}
The ground-state doublet is split by tunnelling into a symmetric (lower) and an
antisymmetric (higher) level. Summing dilute configurations of instantons and
anti-instantons~\cite{coleman1} gives $\Delta E = 2\hbar\mathcal{K}e^{-S_0/\hbar}$
(Supplementary Information, Section~S5). We compare two forms of the prefactor
$\mathcal{K}$. The zero-mode-only form
\begin{equation}\label{eq:coleman}
\Delta E_0 = 2\hbar \omega \sqrt{\frac{S_0}{2\pi \hbar}} \, e^{-S_0/\hbar}
\end{equation}
retains the collective-coordinate Jacobian of the instanton but sets the ratio
of fluctuation determinants to unity, so it is not a consistent one-loop
result. Restoring that ratio, $\det{}'/\det_0 = 1/12$ for this potential, gives
the one-loop splitting
\begin{equation}\label{eq:one_loop}
\Delta E_1 = \sqrt{12}\,\Delta E_0 = 2\hbar \omega \sqrt{\frac{6S_0}{\pi \hbar}} \, e^{-S_0/\hbar}
\end{equation}
\cite{coleman1,escobar_ruiz_2015}, with $\omega = \sqrt{8\lambda a^2/m}$ the
oscillator frequency at the well minimum. Corrections to $\Delta E_1$ form a series in $\hbar/S_0$,
$\Delta E = \Delta E_1\bigl[1 + B_1\hbar/S_0 + \mathcal{O}(\hbar^2/S_0^2)\bigr]$,
with the known two-loop coefficient $B_1=-71/72$~\cite{wohler_shuryak_1994}. Both forms are non-perturbative:
$e^{-S_0/\hbar}$ vanishes faster than any power of $\hbar$.

\subsection{Semiclassical Values at the Benchmark Parameters}\label{sec:sc_prediction}

For the benchmark parameters ($\hbar=1$), $S_0 = \frac{4}{3}a^3\sqrt{2m\lambda}$ and Eq.~\eqref{eq:coleman} give
\begin{align}
S_0 &= \resSZero \\
\omega &= \resOmegaSc \\
\Delta E_0 &\approx \resDESc
\end{align}

The action agrees with the textbook form $S_0 = m^{2}\omega^3/(12\lambda)$~\cite{vainshtein_abc}. At $S_0/\hbar = \resSZero$ the semiclassical expansion is not deep in its asymptotic regime (Section~\ref{sec:determinant_result}).

\section{Lattice Formulation and Extraction of $\Delta E$}\label{sec:lattice}

This section fixes the discretisation, the unary encoding (Supplementary Information, Section~S7) and the ways the splitting is read out.

\subsection{Lattice Discretisation}\label{sec:calibration}
The range $x \in [-X_{\max}, X_{\max}]$ is represented by $N$ sites
\begin{equation}
x_i = -X_{\max} + i \cdot \Delta x, \quad i = 0, 1, \ldots, N-1, \quad \Delta x = \frac{2X_{\max}}{N-1}
\end{equation}

The kinetic operator $-\frac{\hbar^2}{2m}\nabla^2$ is discretised by the three-point stencil $\nabla^2 \psi \approx (\psi_{i+1} - 2\psi_i + \psi_{i-1})/\Delta x^2$, with hopping parameter
\begin{equation}
J = \frac{\hbar^2}{2m \Delta x^2}
\end{equation}
At fixed box, $J$ grows as $N^2$ when the lattice is refined.

\subsection{Unary Encoding}\label{sec:unary_encoding}
In the unary (one-hot) encoding, $N$ qubits represent $N$ positions: $|x_0\rangle=|100\cdots0\rangle$, $|x_1\rangle=|010\cdots0\rangle$, and so on. The physical states form the single-excitation sector, $\hat n=\sum_i(I-Z_i)/2=1$. Hopping and potential terms both conserve $\hat n$, so the exact dynamics stays in this sector without post-selection or penalty terms. A hopping term $-J(|x_i\rangle\langle x_{i+1}| + \text{h.c.})$ maps to $-\frac{J}{2}(X_i X_{i+1} + Y_i Y_{i+1})$, whose exponential is a single Givens rotation costing two CNOTs (Section~\ref{sec:gate_decomposition}).

In the one-excitation sector the Hamiltonian is therefore the $N\times N$
tridiagonal matrix
\begin{equation}
H_{ij}^{\text{unary}} = \begin{cases}
V(x_i) + 2J & i = j \\
-J & |i - j| = 1 \\
0 & \text{otherwise,}
\end{cases}
\label{eq:H_unary}
\end{equation}
where the constant $+2J$ cancels in $\Delta E$.

\subsection{Extraction of the Splitting}\label{sec:methods}

The central observable is the splitting $\Delta E$ between the symmetric ground state $|\psi_0\rangle$ and the antisymmetric first excited state $|\psi_1\rangle$. A particle localised in one well is transferred to the other in the tunnelling time $T_{\text{tunnel}} = \pi\hbar/\Delta E$. We use three methods to obtain $\Delta E$.

\noindent\textbf{Method~I: exact diagonalisation.} Diagonalising the $N\times N$ tridiagonal unary Hamiltonian of Eq.~\eqref{eq:H_unary} gives the spectrum $\{E_n\}$ directly, and $\Delta E = E_1 - E_0$ with it. It is exact to machine precision and is the reference for every other number in this paper (Tier~B). Section~S7.2 of the Supplementary Information constructs the localised states $|L\rangle, |R\rangle$ from its eigenvectors.

\noindent\textbf{Method~II: eigenstate tunnelling dynamics.}\label{sec:method_ii}
Evolving the localised state $|L\rangle = (|\psi_0\rangle+|\psi_1\rangle)/\sqrt2$ gives a survival probability
\begin{equation}\label{eq:survival}
P_L(t) = |\langle L|\psi(t)\rangle|^2 = \cos^2\!\left(\frac{\Delta E\,t}{2\hbar}\right),
\end{equation}
exactly, because $|L\rangle$ has no overlap with higher states. This is the observable a quantum computer measures; Tier~C measures it on a simulated circuit (Section~\ref{sec:tier_c_results}). Evaluated with the exact propagator over $\resEstMethodTwoPeriods$ periods, the peak of the fast Fourier transform (FFT peak) recovers Method~I only to $\resEstMethodTwoFftErrPct\,\%$, the window bias of that estimator, whereas the cosine fit to the same signal recovers it to \resEstMethodTwoFitErr; both estimators are defined at the end of this section.

\noindent\textbf{Method~III: position eigenstate.} A single lattice site needs no eigenstate knowledge and only one $X$ gate, but its weight in the doublet is $\resInitSiteWeightNEight$ at $N=\resNSITES$ and falls roughly like $\Delta x$, to $\resInitSiteWeightLast$ at $N=128$. Over a long window (exact propagator, $t\le60$) the FFT peak of its survival probability recovers Method~I to $\resInitSiteErrLongPct\,\%$; this classical cross-check at $N=\resNSITES$ is the only use of Method~III. On the time grid of the device measurement (Section~\ref{sec:hw_campaign}) the $\cos^2$ fit is biased by $\resInitHwFitErrSitePct\,\%$ for a single-site start, against $\resInitHwFitErrLPct\,\%$ for $|L\rangle$, so the hardware starts from $|L\rangle$. In the unary sector diagonalisation is cheap at any $N$. Where a diagonalisation-free start is needed, a Gaussian packet in the left well (the harmonic ground state, prepared with the same cascade as $|L\rangle$, Section~\ref{sec:trotterization}) keeps a doublet weight of at least $\resInitGaussWeightMin$ for $8\le N\le128$ and a fit bias of $\resInitHwFitErrGaussPct\,\%$ on the device grid; on the
device it reproduces the splitting of $|L\rangle$ to within
$\hwGaussDiffAbsMax$ percentage points (Appendix~\ref{app:hw_repro}).

\noindent\textbf{Frequency estimators.} Since $P_L(t)=\tfrac12\bigl[1+\cos(\Delta E\,t/\hbar)\bigr]$, the splitting is the angular frequency of the sampled survival probability. Two estimators are used. The \emph{FFT peak} subtracts the mean of the samples, pads the series with zeros to eight times its length, computes its discrete Fourier transform by the FFT and refines the position of the largest peak by a parabola through three points. The \emph{cosine fit} is a least-squares fit of $A+B\cos(\omega t+\varphi)$, started from the FFT peak; the device analysis fits the equivalent form $A\cos^2(\Delta E\,t/2\hbar)+C$. Zero padding refines the frequency grid but not the resolution, which is set by the length $T$ of the time window: the transform of a cosine truncated to $T$ is a peak of width $\sim2\pi/T$ at $\Delta E/\hbar$, and its overlap with the mirror peak at $-\Delta E/\hbar$ shifts the maximum. On the exact $N=\resNSITES$ signal the FFT peak is therefore biased by $\resEstFftBiasTierEPct\,\%$ over 1.5 periods and by $\resEstFftBiasSixPeriodsPct\,\%$ over six. The cosine fit has no such bias for a two-level signal; its error stays below \resEstFitErrMax\ for every window from 1.5 to 10 periods. The values of $\Delta E$ extracted from the MPS simulations (Section~\ref{sec:tn_continuum}) and from the device (Section~\ref{sec:hw_campaign}) therefore come from the cosine fit; the FFT peak serves as its starting value and in the classical cross-checks of Methods~II and~III.

\subsection{Model Error and Simulation Error at $N=\resNSITES$}\label{sec:n8_scope}

Table~\ref{tab:convergence} excludes $N=\resNSITES$ from the continuum fit,
yet the gate-level simulations (Tier~C) and the device measurement run at this
size. The two statements concern different errors. A digital simulation of
the lattice Hamiltonian $H(N)$ carries a \emph{model error},
$\Delta E(N)-\Delta E_\infty$, fixed by the choice of $N$ and computable
classically for every $N$, and a \emph{simulation error},
$\Delta E_{\text{sim}}(N)-\Delta E(N)$, from the Trotter step, the circuit
and the device. A quantum computer contributes only the second. Every
simulated splitting in this paper is therefore compared with the exact
splitting of the same lattice (Method~I), never with $\Delta E_\infty$. At
$N=\resNSITES$ the model error is $\resConvRelErrNSites\,\%$, while the
repeated device runs recover the lattice splitting to within
$\hwRepMeanSePct\,\%$ (Eq.~\eqref{eq:hw_repro}); the two do not mix.

The $N=\resNSITES$ model is a coarse but genuine double well. Its minima lie
at $x=\pm\resCoarseXMin$ with $V=\resCoarseVMin$, and the barrier is sampled
only at $x=\pm\resCoarseXBarrier$, where $V=\resCoarseVBarrier$: the lattice
barrier is $\resCoarseBarrierHeight$ high instead of
$\lambda a^4=\resBARRIERLAMBDA$. The lowest doublet is still isolated. The
next level lies $(E_2-E_1)/\Delta E=\resCoarseGapRatio$ splittings above it,
against $\resCoarseGapRatioResolvedMin$ on the resolved grids of the
continuum fit, so the dynamics from $|L\rangle$ is two-mode tunnelling and
$\Delta E$ still falls exponentially with $\sqrt\lambda$
(Section~\ref{sec:instanton_scaling}). What the coarse lattice does not
reproduce is the instanton: its splitting and its barrier dependence are set
by the lattice, not by $S_0$.

Statements about the instanton and the continuum therefore rest on resolved
grids: $N=\resRichardsonWindow$ for the continuum limit and the semiclassical
comparison (Sections~\ref{sec:convergence} and~\ref{sec:semiclassical_comparison}),
and matrix-product-state circuits up to $N=48$, with a model error of
$\resConvRelErrFirstFit\,\%$ (Section~\ref{sec:tn_continuum}). Statements made at
$N=\resNSITES$ concern the simulation and are quoted against the $N=\resNSITES$
lattice. We use $N=\resNSITES$ because current hardware can run it: a full transfer there already takes up to
$\hwTwoQMax$ two-qubit gates, while at $N=\resCsNLast$, next to the first
fitted grid of Table~\ref{tab:convergence}, one tunnelling period needs
$\resCsNeededLast$ Trotter steps against a fidelity budget of
$\resCsBudgetLast$ (Section~\ref{sec:circuit_scaling}).

\section{Exact Diagonalisation Results}\label{sec:exact_diag_results}

We use Method~I to examine the barrier-height dependence and the continuum convergence across
$N=6$ to $N=192$ lattice sites.

\subsection{Barrier-Height Dependence of $\Delta E$}\label{sec:instanton_scaling}

Since $S_0 = \frac{4}{3}a^3\sqrt{2m\lambda}$, instanton theory predicts
\begin{equation}
\ln(\Delta E) = -\frac{S_0}{\hbar} + \text{const} = -\frac{4a^3\sqrt{2m}}{3\hbar} \sqrt{\lambda} + \text{const}
\end{equation}
with predicted slope $-4a^3\sqrt{2m}/(3\hbar)\approx-1.89$ in $\sqrt\lambda$.
At $N=8$, a scan over $\lambda\in[1,10]$ gives a slope of about $-1.02$ over the
upper three quarters of the range. The difference reflects the coarse lattice
and the $\lambda^{1/4}$ dependence of the prefactor. A splitting that falls with
the barrier height is not by itself an instanton signature, since any barrier
suppresses tunnelling; the instanton physics is tested by the continuum
comparison in Section~\ref{sec:semiclassical_comparison}.

\begin{figure}[htbp]
\centering
\includegraphics[width=\textwidth]{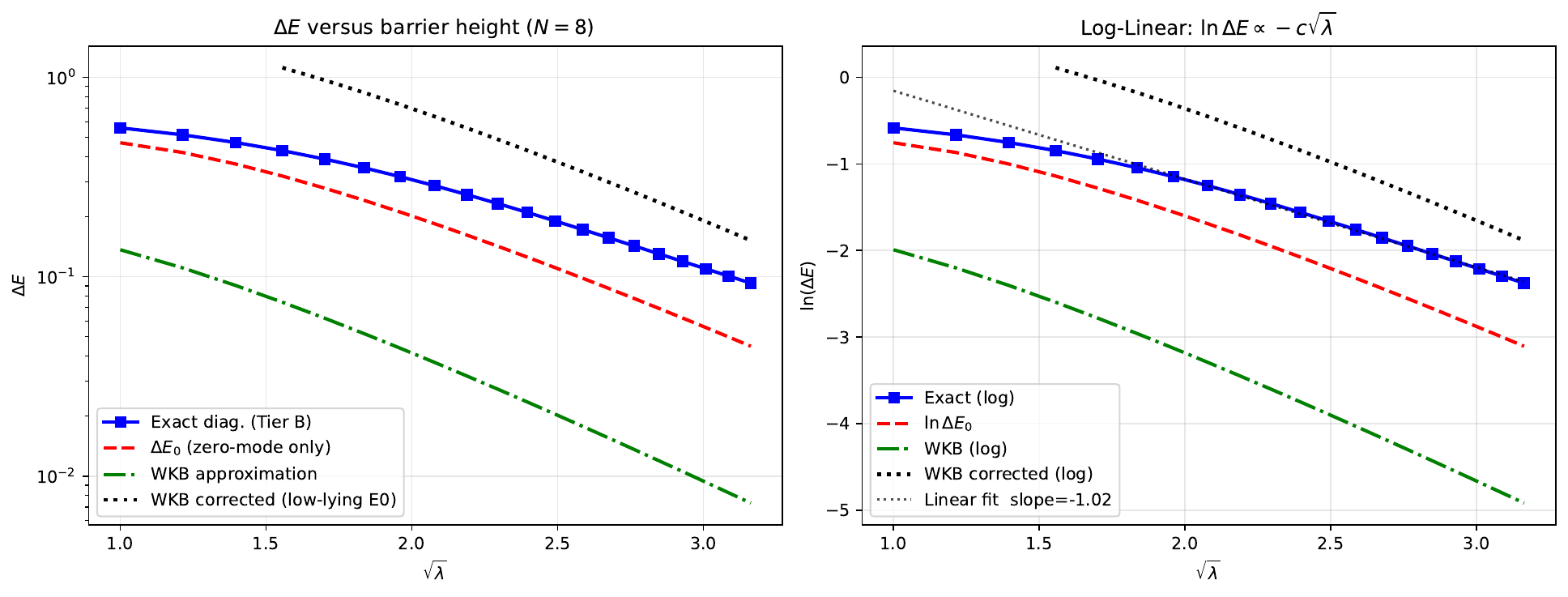}
\caption{Exact unary-Hamiltonian splitting (blue) and the zero-mode-only prediction $\Delta E_0$ (red dashed) versus $\sqrt{\lambda}$ at $N=8$, with a linear fit of $\ln\Delta E$ on the right. The fit gives slope $-1.02$, against the pure-exponential prediction $-1.89$. The WKB curves are discussed in Appendix~\ref{app:determinant_check}.}
\label{fig:instanton_signature}
\end{figure}

\subsection{Continuum Limit}\label{sec:convergence}

\begin{figure}[htbp]
\centering
\includegraphics[width=1.0\textwidth]{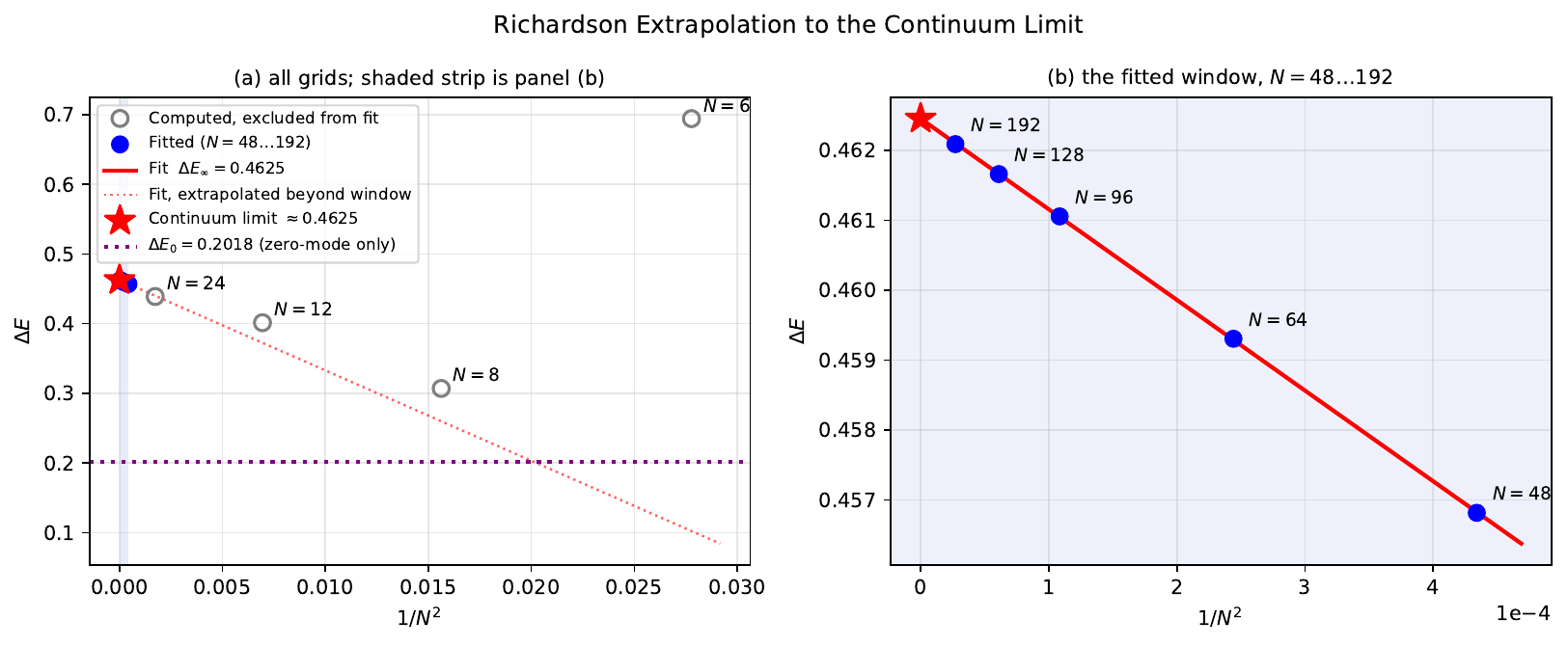}
\caption{Continuum-limit study for the unary Hamiltonian (Tier~B,
$x\in\resXdomain$). \textbf{Left:} Exact splittings versus $1/N^2$; filled blue
points enter the fit ($N=\resRichardsonWindow$), open grey points are excluded,
and the dotted fit extension is illustrative only. The zero-mode-only
prediction $\Delta E_0$ is shown in purple; its missing determinant prefactor is discussed
in Section~\ref{sec:determinant_result}. \textbf{Right:} Magnification of the fit
window and extrapolated value $\Delta E_\infty\approx\resDEInf$ (red star).}
\label{fig:continuum_limit_ext}
\end{figure}

We now vary $N$ to extrapolate to the continuum limit $\Delta E_\infty$ (Figure~\ref{fig:continuum_limit_ext}); since
$N$ is also the qubit number, this fixes the qubit resources. On resolved grids
the finite-difference error is $\mathcal{O}(\Delta x^2)=\mathcal{O}(N^{-2})$,
and we fit
\begin{equation}
\Delta E(N) = \Delta E_\infty + \frac{c_2}{N^2} + \mathcal{O}(N^{-4}),
\end{equation}
taking $\Delta E_\infty$ as the intercept of a linear fit in $1/N^2$. Coarse grids
do not follow this expansion: at $N=8$ on $x\in\resXdomain$, $\Delta E(N)$ is
not yet monotone. On $N=12$--$48$ the $\mathcal{O}(N^{-4})$ term is still
appreciable, and including these grids biases the intercept low by
\resRichardsonWindowLowRel.

We set the fit window against an independent reference: extrapolating directly
in $\Delta x^2$ at fixed box from $\Delta x=\resRichardsonReferenceDx$ gives
$\Delta E=\resDEContinuumRef$. The coarsest window reproducing it to better than
$0.1\%$ is $N=\resRichardsonWindow$, with difference \resRichardsonVsRefRel.
Box half-widths $L=2$--$6$ give a stable result from $L=\resBoxLconverged$;
at the chosen $L=\resXhi$ the residual box dependence is \resBoxRelDev.

\begin{table}[ht]
\centering
\begin{tabular}{|c|c|c|c|l|}
\hline
$N$ & $\Delta x$ & $\Delta E$ & error vs.\ $\Delta E_\infty$ & role \\
\hline
6 & 1.2000 & 0.6942 & $+50.1\%$ & excluded$^{a}$ \\
8$^{\dagger}$ & 0.8571 & 0.3069 & $-33.6\%$ & excluded$^{a}$ \\
\hline
12 & 0.5455 & 0.4013 & $-13.2\%$ & below window$^{b}$ \\
24 & 0.2609 & 0.4389 & $-5.1\%$ & below window$^{b}$ \\
\hline
48 & 0.1277 & 0.4568 & $-1.2\%$ & fitted \\
64 & 0.0952 & 0.4593 & $-0.7\%$ & fitted \\
96 & 0.0632 & 0.4611 & $-0.3\%$ & fitted \\
128 & 0.0472 & 0.4617 & $-0.2\%$ & fitted \\
192 & 0.0314 & 0.4621 & $-0.1\%$ & fitted \\
\hline
\multicolumn{2}{|c|}{$N\to\infty$} & $\approx \resDEInf$ & $c_2 = \resCTwoCThree$ & extrapolated \\
\hline
\end{tabular}
\caption{Convergence of the energy splitting $\Delta E$ with lattice resolution, from Method~I (exact diagonalisation in the unary subspace) at $\lambda=\resBARRIERLAMBDA$, $a=\resMINIMAA$, $m=\resPHYSICALMASS$, $\hbar=1$, $x\in\resXdomain$. The relative error is taken against the Richardson-extrapolated continuum limit $\Delta E_\infty \approx \resDEInf$, fitted over $N=\resRichardsonWindow$. On the fitted window the convergence is monotonic and \emph{from below} ($c_2 = \resCTwoCThree < 0$). The excluded grids are listed because they are the evidence for where the window starts. $^{a}$~The grid misses the minima and samples the barrier only at its flanks, so $\Delta E(N)$ is not yet on the $1/N^2$ branch. $^{b}$~The wells are resolved, but an $O(N^{-4})$ residual remains that a linear fit in $1/N^2$ cannot absorb; fitting $N=12\ldots48$, as done previously, returns a limit \resRichardsonWindowLowRel\ below the value obtained directly at $\Delta x = \resRichardsonReferenceDx$. $^{\dagger}$~The lattice of the gate-level simulations (Tier~C) and of the device measurement. Its own splitting, not $\Delta E_\infty$, is their reference: the discretisation error is a property of the model, not of its simulation (Section~\ref{sec:n8_scope}).}
\label{tab:convergence}
\end{table}

Appendix~\ref{sec:extrapolation_robustness} varies the fit window, form and
points. The largest spread, the relative difference of \resRichardsonVsRefRel\ from
the direct reference, is taken as the uncertainty:
\begin{equation}\label{eq:dEinf_quoted}
\Delta E_\infty = \resDEInfQuoted,
\end{equation}
i.e.\ \resDEInfUncertaintyRel\ relative; the box contributes \resBoxRelDev\ and the eigensolver \resSolverRel.

\label{sec:convergence_projections}
With $c_2\approx\resCTwoCThree$ (verified through $N=\resRichardsonNmax$) the
discretisation error falls below $1\%$ at $N\approx55$ and below $0.2\%$ at
$N\approx130$. Section~\ref{sec:circuit_scaling} projects the circuit depth such
lattices require.

\section{Comparison with the Semiclassical Prediction}\label{sec:semiclassical_comparison}

We compare the exact splitting with the dilute-instanton-gas forms $\Delta E_0$ and $\Delta E_1$ and with WKB~\cite{garg_2000,zinn_justin_QFT}.

\subsection{Ratio of Exact to Semiclassical Splitting}\label{sec:coleman_benchmarks}\label{sec:convergence_semiclassical}

For the standard parameters ($\lambda = \resBARRIERLAMBDA$, $a = \resMINIMAA$,
$m = \resPHYSICALMASS$) the ratio to the zero-mode-only form (Eq.~\eqref{eq:coleman})
is $\Delta E_{\text{exact}} / \Delta E_0 = \resRatioExactSc$ on the
$N=\resNSITES$ lattice and $\Delta E_\infty / \Delta E_0 = \resRatioInfSc$ in
the continuum limit, with $\Delta E_\infty \approx \resDEInf$ and
$\Delta E_0 \approx \resDESc$. This gap is not a quantum correction:
$\Delta E_0$ omits the ratio of fluctuation determinants
(Section~\ref{sec:determinant_result}).

\subsection{Numerical Fluctuation Determinant}\label{sec:determinant_result}

The reduced determinant ratio of the quartic double well is $1/12$, so
restoring it multiplies $\Delta E_0$ by $\sqrt{12}$. Independent Gel'fand--Yaglom
and large-action checks agree to $0.05\%$ (Appendix~\ref{app:determinant_check}).
The one-loop result $\Delta E_1=\resDEScOneLoop$ lies
\resOneLoopOverExactPct\,\% above $\Delta E_\infty=\resDEInf$ (ratio
$\resRatioInfOneLoop$). The ratio $\Delta E_{\text{exact}}/\Delta E_1$
approaches unity from below as $S_0/\hbar$ grows, consistent with the negative
two-loop coefficient $B_1=-71/72$.

\section{Trotterised Circuits and Statevector Validation}\label{sec:quantum_circuits}

This section builds the Trotterised circuits and validates them against exact
diagonalisation. All results in Sections~\ref{sec:trotterization}--\ref{sec:benchmarking}
are Tier~C: Qiskit's noiseless statevector simulator executes the gate sequence
of a quantum processor without shot noise, decoherence or gate errors, which
isolates the Trotter error. Tier~C uses $N=8$ qubits unless stated otherwise.

\subsection{Trotterised Time Evolution}\label{sec:trotterization}

The circuit measures a probability as a function of time, and the splitting is read out of that time series as a frequency. It starts from $|L\rangle = (|\psi_0\rangle+|\psi_1\rangle)/\sqrt2$, a single excitation spread over the left-well sites. The exact propagator $e^{-iHt/\hbar}$ is not available as a gate sequence. The circuit therefore applies the Trotterised propagator $U(\delta t)^n$, $t=n\,\delta t$: $n$ repetitions of a step $U(\delta t)$ that is a product of exponentials of the individual hopping and potential terms of $H$, each of which compiles to a short gate sequence (Eq.~\eqref{eq:trotter_step} below). The observable is the survival probability $P_L(t) = |\langle L|\psi(t)\rangle|^2$, sampled from bitstrings on hardware and computed from the statevector in Tier~C.

\noindent\textbf{Preparation of $|L\rangle$.} The real amplitudes $c_i=\langle x_i|L\rangle$ are computed classically from the two lowest eigenvectors of Method~I (Section~S7.2 of the Supplementary Information). A cascade along the chain prepares $|L\rangle=\sum_i c_i|x_i\rangle$ from $|0\cdots0\rangle$: an $X$ gate puts the excitation on qubit~$0$. Step $k=0,\dots,N-2$ then applies a controlled $R_Y(2\theta_k)$ with control~$k$ and target~$k+1$, followed by a CNOT with control~$k+1$ and target~$k$, which maps $|1_k0_{k+1}\rangle$ to $\cos\theta_k|1_k0_{k+1}\rangle+\sin\theta_k|0_k1_{k+1}\rangle$. With $\cos\theta_k = c_k/r_k$, $r_k=(\sum_{j\ge k}c_j^2)^{1/2}$, each step leaves amplitude $c_k$ on site~$k$ and passes the remainder $r_{k+1}$ to site~$k+1$. The angles depend only on the eigenvector components $\psi_0(x_i)$ and $\psi_1(x_i)$; the eigenvalues $E_0$, $E_1$, and hence $\Delta E$, do not enter the preparation. The splitting enters the signal only through the relative phase that $|\psi_0\rangle$ and $|\psi_1\rangle$ accumulate during the evolution. The cascade uses nearest-neighbour gates only, has depth $\mathcal{O}(N)$ and is exact for real amplitudes up to the sign of $c_{N-1}$, which a final $Z$ gate restores. The device circuit omits that gate; at $N=\resNSITES$, where $c_{N-1}$ is of order $10^{-6}$, this changes $P_L$ by less than $10^{-10}$. Preparation and its inverse cost $\hwTwoQPrep$ two-qubit gates on the device. Tier~E applies the same cascade; Tier~C sets the initial statevector directly.

Only the relative phase $e^{-i(E_1-E_0)t/\hbar}$ survives in $P_L$ (Eq.~\eqref{eq:survival}), so the accuracy of $\Delta E$ is that of an accumulated phase. The Trotter error enters as phase drift over many steps; the step size $\delta t$, not the gate count, sets the accuracy, and the extracted $\Delta E$ inherits the $\mathcal{O}(\delta t^2)$ convergence of the second-order formula. The splitting is extracted with the cosine fit of Section~\ref{sec:methods}, which is accurate to better than \resEstFitErrMax; the FFT peak would carry a window bias of several percent.

The unary Hamiltonian of Eq.~\eqref{eq:H_unary} splits into hopping and potential terms, $H=H_T+H_V$, with
\begin{equation}
H_T=-J\sum_{i=0}^{N-2}\left(|x_i\rangle\langle x_{i+1}|+\text{h.c.}\right),
\qquad
H_V=\sum_{i=0}^{N-1}\bigl[V(x_i)+2J\bigr]|x_i\rangle\langle x_i|.
\end{equation}
The second-order Suzuki--Trotter decomposition is
\begin{equation}\label{eq:trotter_step}
e^{-iHt/\hbar} \approx \left( e^{-iH_T \delta t/(2\hbar)} \, e^{-iH_V \delta t/\hbar} \, e^{-iH_T \delta t/(2\hbar)} \right)^{n}
\end{equation}
where $t = n \cdot \delta t$, the bracket is the step $U(\delta t)$ and the error per step is $\mathcal{O}(\delta t^3)$. Qiskit's \texttt{SuzukiTrotter} applies the formula to the individual Pauli terms, so $e^{-iH_T\delta t/(2\hbar)}$ is itself split bond by bond, sweeping the bonds forward and back. The device circuit orders the same bond exponentials in odd and even layers, $O/2\,E/2\,V\,E/2\,O/2$ per step. For both bond-wise formulas the admissible step shrinks roughly as $1/N^2$, because $J\propto N^2$; a splitting that keeps $H_T$ whole does not (Section~\ref{sec:tn_dt_scaling}).

\subsection{Gate Decomposition and Resource Estimates}\label{sec:gate_decomposition}

Synthesised as a single Givens rotation, each hopping term $e^{-i\theta(X_i X_{i+1} + Y_i Y_{i+1})}$ needs 2 CNOT gates plus single-qubit rotations, so the minimal CNOT count per Trotter step is $2(N-1)$. Qiskit's \texttt{PauliEvolutionGate} with \texttt{SuzukiTrotter(order=2)} synthesis instead emits the $XX$ and $YY$ rotations separately, at 2 CNOTs each.

Transpiled to the basis $\{\texttt{cx}, \texttt{rz}, \texttt{sx}, \texttt{x}, \texttt{id}\}$ at optimisation level~3, one Trotter step at $N=8$ ($\delta t = 0.1$) has depth 103, 26 CNOTs (against $2(N-1) = 14$) and 178 gates in total. The excess comes from routing and from emitting $XX$ and $YY$ separately (Supplementary Information, Section~S8). Four per-step counts at $N=8$ appear in this paper: $56$ ($XX$ and $YY$ separate, level~1), $26$ (the same, level~3, as above), $28$ (one Givens rotation per bond, level~1, Supplementary Information, Section~S8) and $\hwTwoQPerStep$ (Givens, level~3, the minimum, reached in the device runs of Section~\ref{sec:hw_campaign}).

\subsection{Tier C Simulation Results}\label{sec:tier_c_results}

\begin{figure}[htbp]
\centering
\includegraphics[width=0.85\textwidth]{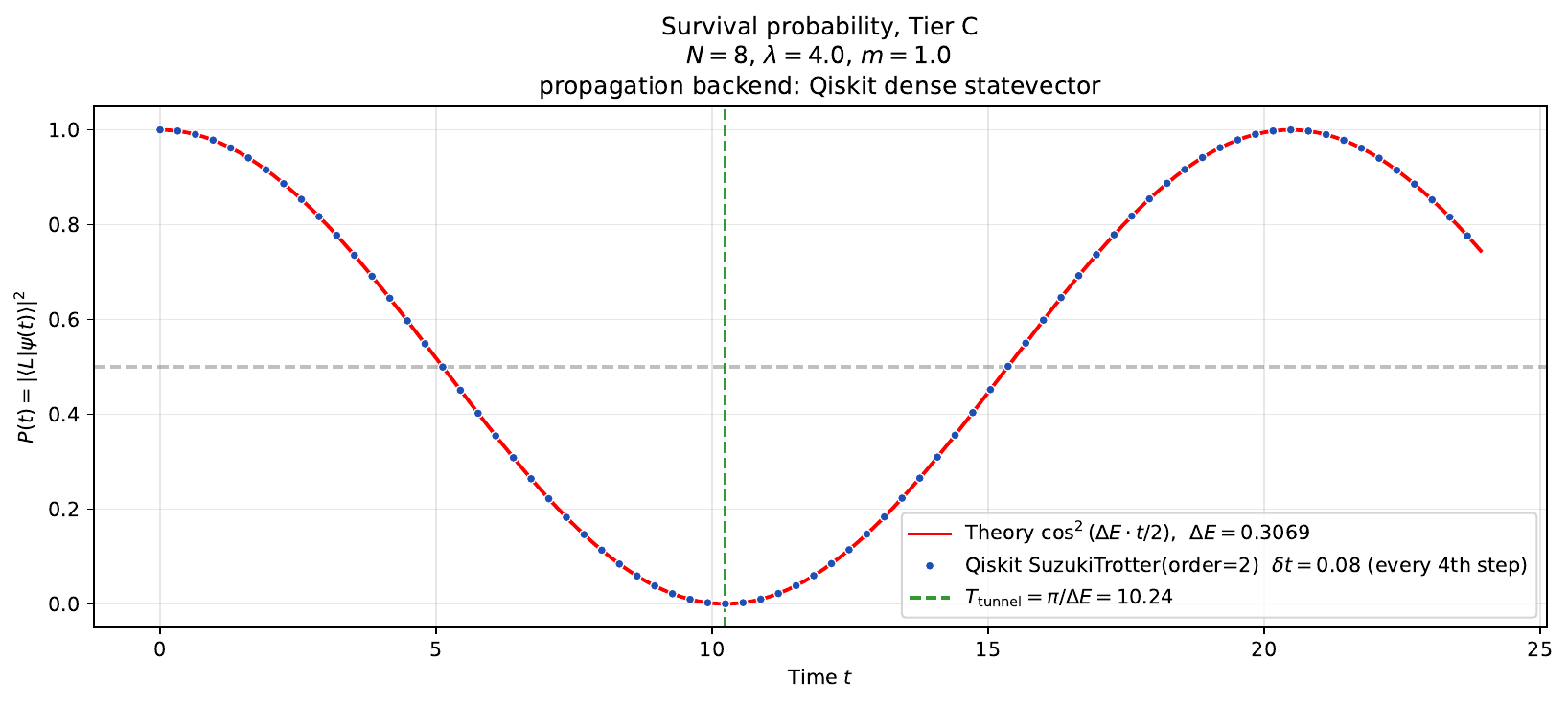}
\caption{Survival probability $P(t) = |\langle L|\psi(t)\rangle|^2$ of the Trotterised circuit (Tier~C, blue dots, every fourth step shown) at $N=\resNSITES$ and the exact $\cos^2(\Delta E\,t/2)$ with $\Delta E = \resDeltaE$ (red). The dashed line marks $T_{\text{tunnel}} = \pi/\Delta E \approx \resTtunnel$.}
\label{fig:tunneling}
\end{figure}

We validate the Trotterised circuit on the noiseless statevector simulator at $N=\resNSITES$ in three respects, taking the exact splitting of this lattice as the reference (Section~\ref{sec:n8_scope}).
First, the survival probability reproduces the exact oscillation of
Eq.~\eqref{eq:survival}, and the splitting recovered from it agrees with
diagonalisation (Figure~\ref{fig:tunneling}). Second, halving $\delta t$
reduces the error of $\Delta E$ as $\mathcal{O}(\delta t^2)$. Third, the
excitation number stays within $10^{-12}$ of unity.

Repeating the $\lambda$ scan of Figure~\ref{fig:instanton_signature} with the
Trotterised circuit reproduces the slope of the lattice splitting
(Figure~\ref{fig:hw_signature}, together with the device measurement of the same
scan).

\subsection{Trotter Error and Subspace Leakage}\label{sec:benchmarking}

The phase fidelity of $P_L(t)$, the mean of $1-|P_L^{\text{Trotter}}-P_L^{\text{exact}}|$
over two tunnelling periods, is $>0.994$ at $\delta t=0.1$ and $\resFidTwoTwob$ at
$\delta t=0.05$. Leakage out of the one-excitation subspace stays below $10^{-12}$,
so the synthesised gates respect the constraint. At large barriers
($\lambda\gtrsim20$) the splitting becomes too small to extract from a time
series in double precision, so the determinant study of
Section~\ref{sec:determinant_result} uses diagonalisation.

\section{Circuit Structure and Classical Simulation}\label{sec:tier_e}

The classical baseline for this model is the $N$-dimensional single-excitation dynamics, which costs $\mathcal{O}(N)$ per time step and stays cheap at any $N$. A tensor-network contraction is therefore not needed to reach large $N$. Its role is to execute the synthesised gate list itself, gate ordering and Trotter error included, at sizes beyond a dense statevector, and to measure the bond dimension that gate list requires. Tier~E contracts the Suzuki--Trotter circuit that Qiskit synthesises as a matrix product state~\cite{schollwoeck_2011} using quimb~\cite{gray_quimb}. (Qiskit Aer~0.17.2 rejects circuits with more than 63 qubits in its default configuration.)

\subsection{Schmidt-Rank Bound of the Trotter Circuit}\label{sec:tn_chi}

A one-excitation state is a weighted $W$ state with Schmidt rank at most $2$ across any cut~\cite{raveh_nepomechie_2024}. However, \texttt{SuzukiTrotter} splits the hopping term $XX+YY$ into separate \texttt{rxx} and \texttt{ryy} rotations, neither of which conserves excitation number, so the state leaves the unary sector between the two gates of a pair. The bond dimension needed without truncation is four in every case we tested (Table~\ref{tab:tn_chi}), and Lemma~\ref{prop:chi4} shows that it cannot exceed four. Applying each bond's $XX+YY$ rotation as a single Givens gate~\cite{kivlichan_2018,arrazola_2022}, as on hardware, makes every gate boundary the end of a pair, so the rank never exceeds two.

\begin{lemma}\label{prop:chi4}
Consider an open chain of $N$ qubits with Hamiltonian
\[
H=-\frac{J}{2}\sum_{b=0}^{N-2}
\left(X_bX_{b+1}+Y_bY_{b+1}\right)
+\sum_{j=0}^{N-1}V_j\frac{I-Z_j}{2}.
\]
Here $J$ and $V_j$ are real.
Let the initial state be any normalized state in the one-excitation sector.
Assume each Trotter step consists of on-site potential unitaries and, for
each bond $b$, the consecutive gates
\[
U_{b,XX}(\theta_b)=e^{-i\theta_bX_bX_{b+1}},
\qquad
U_{b,YY}(\theta_b)=e^{-i\theta_bY_bY_{b+1}},
\]
with the same angle $\theta_b$ in the pair. The order of the bond pairs may
be arbitrary, the two gates of a pair may come in either order, and a bond may
occur several times per step with different angles (as in a symmetric
formula); what is required is that every occurrence is a matched-angle pair
not interleaved with other gates. Then, at every gate boundary, the Schmidt rank across every
contiguous bipartition is at most $4$. At the end of each bond pair, and
hence at the end of each complete Trotter step, the state is in the
one-excitation sector and has Schmidt rank at most $2$ across every cut.
These bounds are independent of $N$, the cut, and the angles.
\end{lemma}

\begin{proof}
Fix a contiguous cut $A|B$. Any one-excitation state admits a decomposition
\[
|\psi\rangle
=|\psi_A\rangle\otimes|0_B\rangle
+|0_A\rangle\otimes|\psi_B\rangle,
\]
where each nonzero component has its excitation in the indicated subsystem.
It is therefore a sum of at most two product vectors, and its Schmidt rank
across the cut is at most $2$.

We use the following rank bound. If an operator crossing a bipartition has
an operator-Schmidt decomposition with $r$ product terms,
$O=\sum_{\alpha=1}^{r} A_\alpha\otimes B_\alpha$, then
\[
\operatorname{SR}(O|\phi\rangle)
\leq r\,\operatorname{SR}(|\phi\rangle).
\]
Indeed, expanding $|\phi\rangle=\sum_{k=1}^{s}|a_k\rangle|b_k\rangle$,
where $s=\operatorname{SR}(|\phi\rangle)$, gives an expansion of
$O|\phi\rangle$ with at most $rs$ product terms. Gates supported entirely
in $A$ or in $B$ preserve Schmidt rank.

For the fixed cut, nearest-neighbour geometry implies that at most one bond
straddles it. Gates on all other bonds are local with respect to this cut.
If the acting bond straddles it, each of its two gates has operator-Schmidt
rank at most $2$, since, for example,
\[
e^{-i\theta X\otimes X}
=\cos\theta\,I\otimes I-i\sin\theta\,X\otimes X,
\]
and the same expansion holds with $Y$ in place of $X$. At the start of a
bond pair, the state is in the one-excitation sector and hence has Schmidt
rank at most $2$. The first gate can therefore increase the rank to at most
$4$.

The two generators commute, and the matched-angle pair satisfies
\[
e^{-i\theta XX}e^{-i\theta YY}
=e^{-i\theta(XX+YY)}.
\]
Define the raising and lowering operators by
$\sigma^\pm=(X\pm iY)/2$. Then
$XX+YY=2(\sigma^+\otimes\sigma^-+\sigma^-\otimes\sigma^+)$, so this pair
preserves total excitation number. The
on-site potential unitaries do so as well. Thus, after each pair, the state returns to the one-excitation
sector, where its Schmidt rank across every cut is at most $2$. Applying
this reasoning successively to all pairs proves both bounds at every gate
boundary and at the end of each Trotter step. The bounds do not depend on
the chain length, cut position, or gate angles.
\end{proof}

The matched-angle adjacency condition is essential. If the synthesis emits
all $XX$ rotations before all $YY$ rotations, the state need not return to the
one-excitation sector within a step; the rank then reaches $8$ in the first step
and, at $N=10$, the maximal $2^{N/2}=32$ by the second.

A direct check confirms the bound: applying the synthesised gate list
(Qiskit~2.3.0) to a dense statevector gives a maximum rank of
\resRankMaxIntermediate\ after any gate and \resRankMaxStepEnd\ at the end of
each step, for $N=6$, $8$ and $10$ over three steps. The bound is attained only
on the cut the acting gate straddles, so $\chi=4$ involves no truncation at all.

\begin{table}[ht]
\centering
\begin{tabular}{|c|c|c|c|c|}
\hline
$N$ & $\chi=1$ & $\chi=2$ & $\chi=4$ & $\chi=8$ \\
\hline
16 & $7.5\times10^{-1}$ & $2.2\times10^{-3}$ & $2.6\times10^{-7}$ & $2.6\times10^{-7}$ \\
32 & $9.9\times10^{-1}$ & $8.4\times10^{-2}$ & $8.5\times10^{-6}$ & $8.5\times10^{-6}$ \\
64 & $1.0\times10^{0}$ & $8.8\times10^{-1}$ & $1.6\times10^{-4}$ & $1.6\times10^{-4}$ \\
\hline
\end{tabular}
\caption{Maximum deviation $\max|\Delta P|$ of the survival probability from exact unary propagation, as a function of MPS bond dimension ($\delta t = 0.002$, 150 steps). At every $N$, $\chi=4$ and $\chi=8$ agree to machine precision, while $\chi=2$ fails and degrades rapidly with $N$. The residual at $\chi\geq4$ is Trotter error, not truncation error.}
\label{tab:tn_chi}
\end{table}

The bound is independent of $N$, so the unary-encoded double well has no bond-dimension bottleneck and cannot test the scaling of tensor-network methods. That requires a system whose entanglement grows after a quench~\cite{calabrese_cardy_2005}, such as an interacting or field-theoretic extension (Section~\ref{sec:future_outlook}); Tier~F shows the contrast on an Ising chain (Appendix~\ref{sec:tier_f}). Non-stabilizerness (magic), which governs the cost of stabilizer-based simulation, grows only logarithmically with $N$ along a tunnelling period while the bond dimension stays fixed (Appendix~\ref{sec:tn_magic}).

\subsection{Trotter Step versus System Size}\label{sec:tn_dt_scaling}

Since $J\propto N^2$ and a single bond term acts on a smooth state with
strength $\sim J$, a product formula that exponentiates every bond separately
needs $J\delta t/\hbar$ fixed, so its admissible step $\delta t^{*}$ shrinks
roughly as $N^{-2}$. The kinetic operator as a whole acts on smooth states with
strength $\mathcal{O}(1)$, and the Strang splitting $e^{-iH_V\delta t/2\hbar}e^{-iH_T\delta t/\hbar}e^{-iH_V\delta t/2\hbar}$
has an $\mathcal{O}(\delta t^2)$ error on smooth solutions uniformly in the grid
spacing~\cite{jahnke_lubich_2000}; commutator bounds draw the same distinction for
product formulas in general~\cite{childs_2021,childs_su_2019}, and error bounds in
the vector norm make the same point for unbounded and discretised operators
acting on regular states~\cite{an_fang_lin_2021,burgarth_2024}. It is the
split-operator method~\cite{wiesner_1996,zalka_1998,kassal_2008,childs_realspace_2022}.

\begin{figure}[htbp]
\centering
\includegraphics[width=\textwidth]{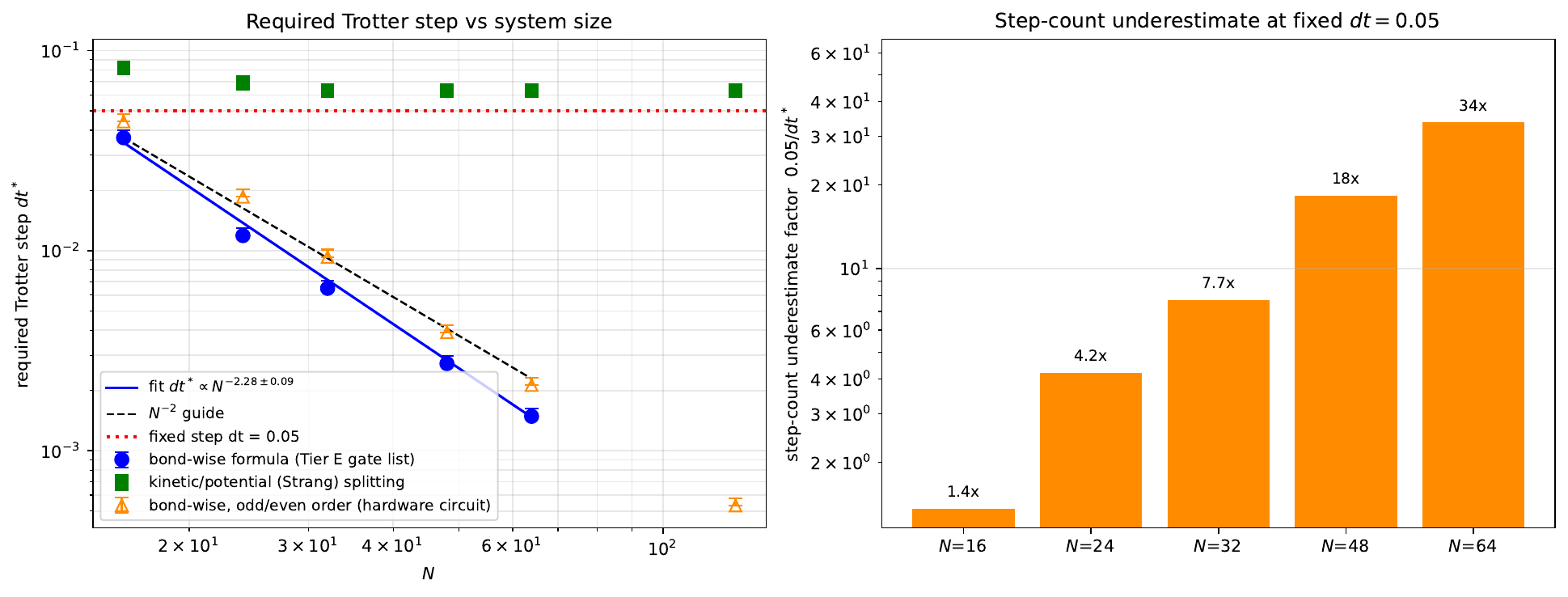}
\caption{\textbf{Left:} $\delta t^{*}$ versus $N$ for the bond-wise gate list
(blue, with the bracket to the next step tested), its fitted power law and an
$N^{-2}$ guide, for the odd/even ordering of the device circuit (open
triangles), which scales alike with a larger step, and for the
kinetic/potential (Strang) splitting (green), whose step does not shrink with
$N$. \textbf{Right:} factor by which a projection that
costs every $N$ at $\delta t=0.05$ understates the number of bond-wise steps.}
\label{fig:tn_dt_scaling}
\end{figure}

We measure $\delta t^{*}$ as the largest step that keeps
$\max|\Delta P|<10^{-3}$ over a unit-time window, for
$N\in\{16,24,32,48,64\}$, by bisection in $\log\delta t$ to a resolution of
$2^{1/8}$ (Figure~\ref{fig:tn_dt_scaling}). For the bond-wise gate list,
fitted through the bracket midpoints with an error that includes their width,
\begin{equation}
\delta t^{*}(N) \propto N^{\resTierEDtStarExponentRefined\pm\resTierEDtStarExponentRefinedErr}
\end{equation}
over all five sizes; for $N\ge24$ the exponent is
$\resTierEDtStarExponentTail\pm\resTierEDtStarExponentTailErr$. To test the
dependence on the criterion we repeat the measurement in the single-excitation
sector for thresholds $10^{-2}$, $10^{-3}$ and $10^{-4}$ over one time unit and
for $10^{-3}$ over one full oscillation period. Over all five sizes the exponent
of the bond-wise sweep then lies between $\resCritBondExpMin$ and
$\resCritBondExpMax$; the steepening comes from $N=16$ and depends on the
criterion. For $N\ge24$ it lies between $\resCritBondTailMin$ and
$\resCritBondTailMax$ under every criterion, for the sweep and for the odd/even
ordering alike. The $N^{-2}$ scaling is therefore robust, its excess over $-2$
is not. One period costs about $N^{\resTierECostExponentRefined}$ two-qubit
gates under the Tier~E criterion. A fourth-order Suzuki formula built from the same bond-wise steps
relaxes $\delta t^{*}$ by a factor
\resSplitOrderFourRelaxMin--\resSplitOrderFourRelaxMax, but needs five
second-order stages per step, so the gate count per unit time improves only by a
factor \resSplitOrderFourNetMin--\resSplitOrderFourNetMax.

For the Strang splitting, evaluated in the single-excitation sector under the
same criterion, $\delta t^{*}$ stays between
$\resSplitTvDtMin$ and $\resSplitTvDtMax$ for $N=16$--$128$ (exponent
$\resSplitTvExponent$ for $N\ge24$); the same sector calculation reproduces the
bond-wise values of the gate list to within \resSplitBondMpsMaxDevSteps\
bisection steps. Its kinetic exponential is an $N\times N$ unitary on the sector,
which a mesh of $N(N-1)/2$ nearest-neighbour Givens rotations implements exactly
in depth $\mathcal{O}(N)$~\cite{clements_2016,kivlichan_2018}. A step then costs
$N(N-1)$ CNOTs, but the step does not shrink.

Costs must pair each step with the gate count of its own circuit. The sweep
ordering of the Tier~E gate list needs $2(2N-3)$ CNOTs per step after merging at
the step boundary (26 at $N=8$). The odd/even ordering of the device circuit
merges into one two-qubit block per bond, $2(N-1)$ CNOTs (14 at $N=8$); it is
bond-wise as well and scales alike, with exponent
$\resSplitEoExponent$ for $N\ge24$, with a step larger by a factor
\resSplitEoDtRatioMin--\resSplitEoDtRatioMax. Per unit time the odd/even circuit
needs $\resSplitEoCostRatioSixFour$ times the two-qubit gates of the Strang
splitting at $N=64$ and $\resSplitEoCostRatioOneTwoEight$ times at $N=128$, the
two being equal near $N\approx\resSplitEoCrossoverN$ (near
$N\approx\resSplitSweepCrossoverN$ for the sweep ordering). The $N^{3.3}$ cost
belongs to splitting the kinetic term bond by bond; with the kinetic term kept
whole, the cost per period grows as $N^2$. Both are product-formula results, not
bounds on post-Trotter methods such as qubitization~\cite{low_chuang_2019} or
truncated-Taylor simulation~\cite{berry_2015}.

Costing every $N$ at a fixed step $\delta t=0.05$ understates the CNOT count
of the bond-wise formula by more than an order of magnitude at $N=64$.
Section~\ref{sec:circuit_scaling} uses the bond-wise formula run on hardware.

\subsection{Continuum Limit with Matrix-Product States}\label{sec:tn_continuum}

\begin{figure}[htbp]
\centering
\includegraphics[width=0.85\textwidth]{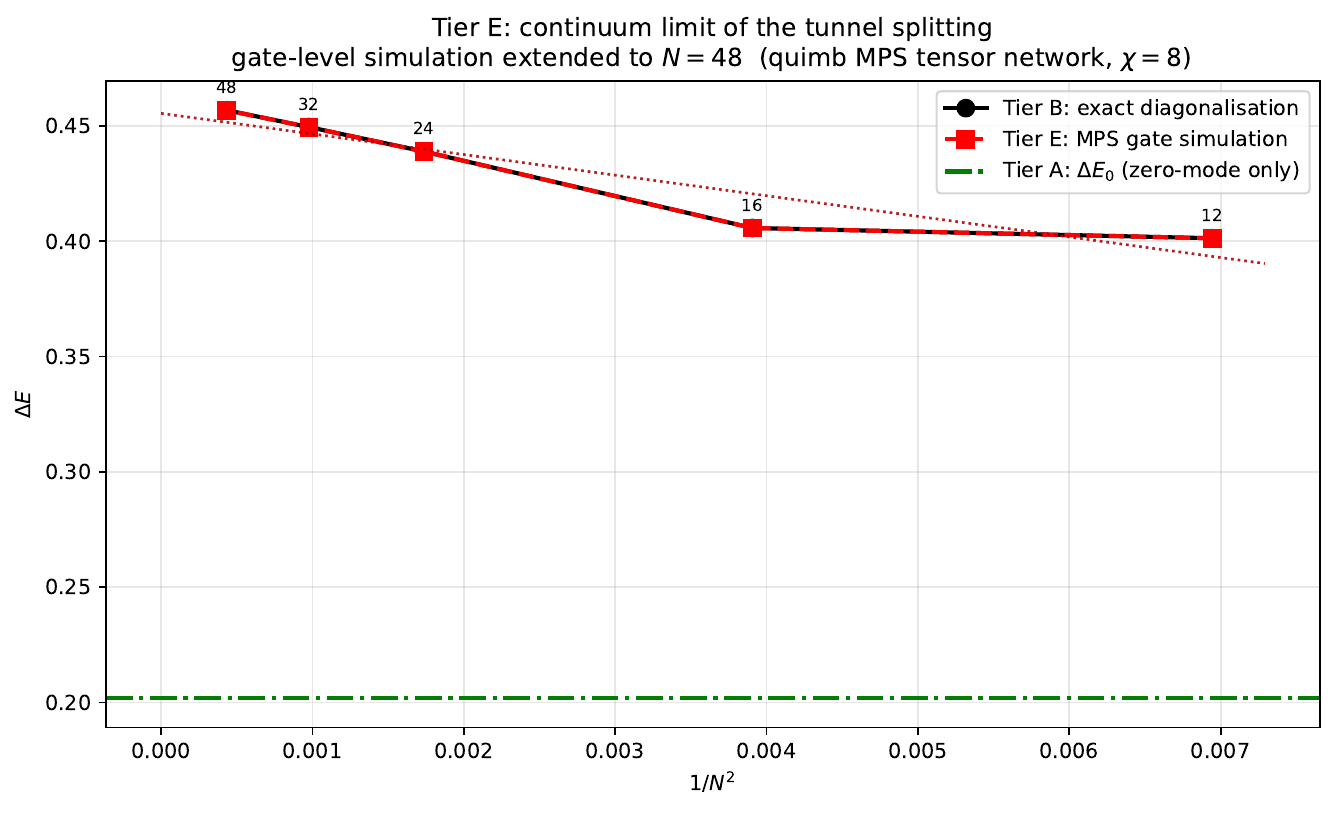}
\caption{Finite-window $1/N^2$ extrapolation of the tunnel splitting from the gate-level MPS simulation (Tier~E, red) and exact diagonalisation (Tier~B, black), with the zero-mode-only prediction $\Delta E_0$ (Tier~A, green). Neither extrapolation shown is the continuum estimate quoted in Section~\ref{sec:convergence}.}
\label{fig:tn_continuum}
\end{figure}

We extend the gate-level circuit to $N=48$. The splitting is extracted with
the cosine fit of Section~\ref{sec:methods}; over the 1.5 simulated periods the
FFT peak would be biased by $\resEstFftBiasTierEPct\,\%$. After removing the
$\mathcal{O}(\delta t^2)$ Trotter error by extrapolation in $\delta t$, the
MPS and exact diagonalisation results agree to at worst
\resTierEWorstRel\ over $N\in\{12,16,24,32,48\}$. Their finite-window
$1/N^2$ extrapolations differ by \resTierEContinuumRel\ (Figure~\ref{fig:tn_continuum}),
but neither is the continuum limit: the accessible window is the coarse one
rejected in Section~\ref{sec:convergence}, and both lie
\resRichardsonWindowLowRel\ below $\Delta E_\infty$. Tier~E validates the
circuit at larger sizes; it does not provide an independent continuum estimate.

\section{Hardware Projections and Measurement}\label{sec:outlook}

This section projects the hardware cost from gate counts and a per-gate error
model (Section~\ref{sec:circuit_scaling}; a calibrated-noise study is in
Section~S8 of the Supplementary Information) and reports the device measurement
(Section~\ref{sec:hw_campaign}).

\subsection{Gate-Count Projections}\label{sec:circuit_scaling}

Each Trotter step of the device circuit needs $2(N-1)$ CNOTs, one Givens rotation per bond. At $N = 100$ a single step uses $\sim 200$ CNOTs and, at a CX error rate of $2\times10^{-3}$, retains a fidelity of only $(0.998)^{200} \approx 0.67$. Table~\ref{tab:circuit_scaling} compares, for each $N$, the Trotter steps needed to span one tunnelling period $T_{\text{tunnel}}=\pi\hbar/\Delta E$ at the step measured for this circuit with the steps that fit into this fidelity budget.

\begin{table}[ht]
\centering
\setlength{\tabcolsep}{4pt}
\begin{tabular}{|c|c|c|c|c|c|c|}
\hline
\textbf{Qubits} & \textbf{Required} & \textbf{CNOTs /} & \textbf{Fidelity /} & \textbf{Steps in} & \textbf{Steps} & \textbf{Total CNOTs} \\
\textbf{($N$)} & \textbf{step $\delta t^{*}$}$^{\,a}$ & \textbf{Trotter step} & \textbf{step}$^{\,b}$ & \textbf{budget}$^{\,c}$ & \textbf{needed}$^{\,d}$ & \textbf{needed} \\
\hline
12 & 0.0573 & 22 & 0.96 & 15 & 137 & $3.0\times10^{3}$ \\
16 & 0.0442 & 30 & 0.94 & 11 & 175 & $5.3\times10^{3}$ \\
20 & 0.0287 & 38 & 0.93 & 9 & 256 & $9.7\times10^{3}$ \\
25 & 0.0170 & 48 & 0.91 & 7 & 418 & $2.0\times10^{4}$ \\
32 & 0.0093 & 62 & 0.88 & 5 & 752 & $4.7\times10^{4}$ \\
50 & 0.0036 & 98 & 0.82 & 3 & 1918 & $1.9\times10^{5}$ \\
\hline
\end{tabular}
\caption{Projected quantum circuit resources for intermediate system sizes, for the bond-wise second-order formula. $^a$~Measured for the odd/even bond ordering of the device circuit on the Tier~E criterion $\max|\Delta P|<10^{-3}$ over one time unit (Section~\ref{sec:tn_dt_scaling}); the lower end of a bisection bracket of width $2^{1/8}$, so conservative by at most 9\%. $^b$~$2(N-1)$ CNOTs per step, the count this ordering reaches with one Givens rotation per bond, as in the device runs of Section~\ref{sec:hw_campaign}, at a CX error rate of $2\times10^{-3}$, the scale of the measured medians. $^c$~Trotter steps before cumulative fidelity falls below $0.50$. $^d$~Steps to span one tunnelling period $T_{\text{tunnel}}=\pi\hbar/\Delta E$ at the required step size. All entries are Tier~D projections from a depolarising, per-gate error model with no crosstalk, idling or readout error and no error mitigation.}
\label{tab:circuit_scaling}
\end{table}

At $N=\resCsNFirst$ the budget permits $\resCsBudgetFirst$ Trotter steps while $\resCsNeededFirst$ are needed for one period; at $N=\resCsNLast$ it permits $\resCsBudgetLast$ against $\resCsNeededLast$. The CNOT count per step grows like $N$ and the admissible step shrinks roughly like $N^{-2}$ (Section~\ref{sec:tn_dt_scaling}), so the total grows roughly as $N^{3}$. Keeping the kinetic term whole lowers the exponent to two but costs more per unit time below $N\approx\resSplitEoCrossoverN$, so it does not change the table.

The projection rests on a per-CX depolarising
model that omits crosstalk, coherent and drifting errors, idling decoherence,
state-preparation and measurement errors, and error mitigation; connectivity
enters only through the transpiled gate counts, and
$p_{\text{CX}}=2\times10^{-3}$ is the scale of the measured medians. Taking the
lower end of the bisection bracket makes the step counts conservative by at most
9\%; imposing the criterion over one time unit rather than a full period makes
them optimistic.

Costing every $N$ at a fixed $\delta t = 0.05$ would understate the step count by up to $\resCsFixedStepFactorLast\times$ at $N=\resCsNLast$, and the ``steps in budget'' column alone counts only how many steps fit before decoherence, not how many are needed. A full tunnelling oscillation at the projection's accuracy ($10^{-3}$ over one time unit) is therefore out of reach at every $N$ considered at $p_{\text{CX}} = 2\times10^{-3}$; it requires an error rate one to three orders of magnitude lower, or early fault tolerance. A fourth-order formula only shifts the problem (Section~\ref{sec:tn_dt_scaling}). The compilation and noise study of Section~S8 of the Supplementary Information reaches the same conclusion.

What remains accessible is the short-time decay of the survival probability, $P(t)\simeq 1-(\Delta E\,t/2)^2$, which needs $\mathcal{O}(10)$ rather than $\mathcal{O}(10^2)$ steps, and a full transfer at a looser, separately bounded step. The device measurement below takes the second route at $N=\hwNSites$. At $N=12$ no scanned step keeps the noiseless circuit within $\hwScanTol$ of the exact evolution, and already the finest, $\delta t=\hwScanTwelveDtMin$, needs $\hwScanTwelveTwoQ$ two-qubit gates.

\subsection{Device Measurement}\label{sec:hw_campaign}

A device measurement is not reproducible, since a later run meets a different
calibration; only the analysis of the frozen data is, and the test suite re-runs
it. The circuit is the odd/even Givens circuit of
Section~\ref{sec:tn_dt_scaling}, executed on \hwBackend, a Heron-generation
processor with median two-qubit error $\hwTwoQError\times10^{-3}$ at the time of
the reference run. We measure the $N=\hwNSites$ lattice, whose exact splitting is the reference
(Section~\ref{sec:n8_scope}), at $\delta t = \hwDt$ over
$\hwPoints$ time points out to $t = \hwTmax$, with $\hwShots$ shots per
circuit; Table~\ref{tab:hw_methods} collects the protocol.

\begin{figure}[htbp]
\centering
\includegraphics[width=\textwidth]{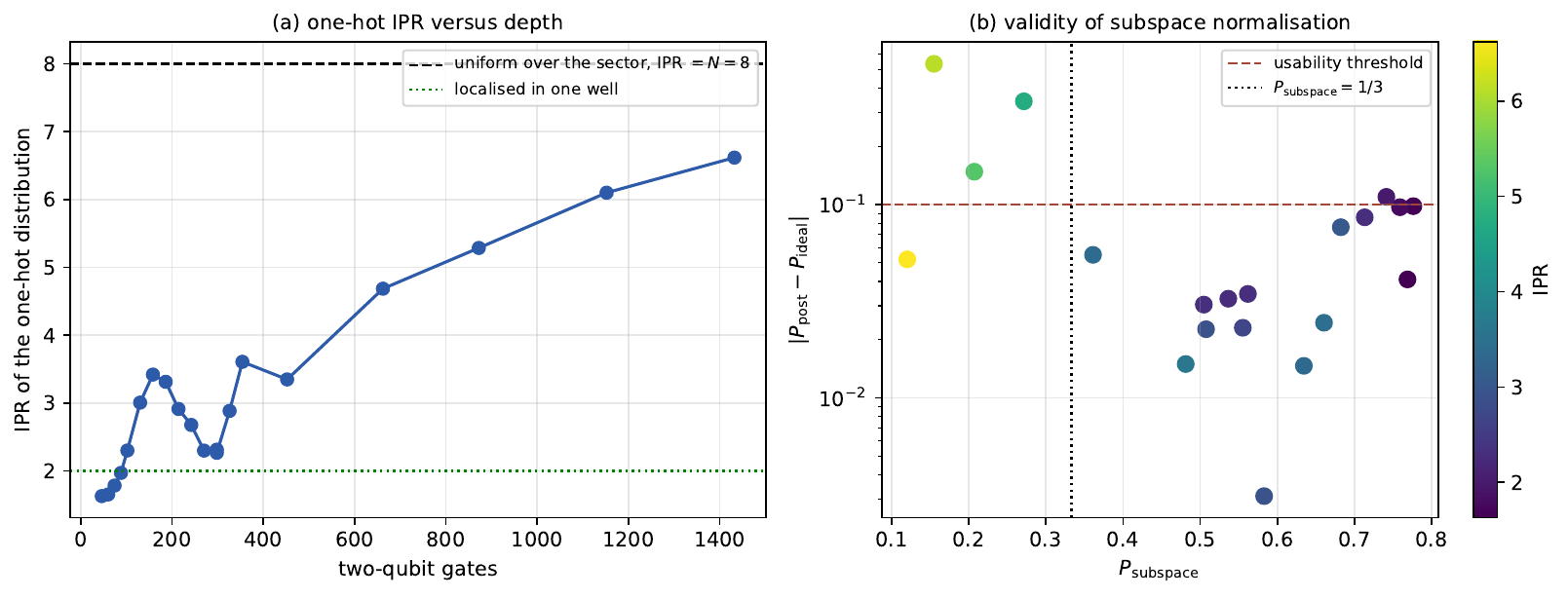}
\caption{\textbf{Validity limit of subspace normalisation}
(\hwBackend; external to the pipeline). \textbf{(a)} Inverse participation
ratio of the one-excitation distribution against circuit depth. Up to about
$\hwValCohTwoQ$ two-qubit gates it oscillates with the tunnelling; beyond, it
rises monotonically toward the uniform value $N$. \textbf{(b)} Deviation of the
normalised estimate from the noiseless circuit against the surviving
subspace population, coloured by participation ratio. The crossover is near
$P_{\text{subspace}} = 1/3$. The scatter within the usable region reflects
where each point sits on $\cos^2$, not the quality of the correction.}
\label{fig:hw_validity}
\end{figure}

Each time point costs two circuits. Circuit~A prepares $|L\rangle$, applies
the Trotter evolution and the inverse preparation, and estimates the survival
probability $P_{\text{raw}}$ as the frequency of the all-zero outcome.
Circuit~B omits the inverse preparation and estimates the population
$P_{\text{subspace}}$ of the single-excitation sector as the frequency of
one-hot outcomes. The estimator is the ratio
$P_{\text{post}} = P_{\text{raw}}/P_{\text{subspace}}$, a symmetry-based
normalisation in the spirit of symmetry
verification~\cite{bonet_monroig_2018,mcardle_2019} rather than post-selection
on the shots of circuit~A; we call it \emph{subspace normalisation}. It is
unbiased if the loss of overlap in~A equals the leakage measured in~B, which we
test only indirectly: against the noiseless circuit, by the validity limit of
Figure~\ref{fig:hw_validity} and by the layout check below. $P_{\text{subspace}}$
stays above $\hwKingPsubMin$ in every run on \hwBackend. Since $|L\rangle$ is
prepared from the eigenvectors of Method~I, the circuit measures the phase of a
known two-level superposition; a single-site start would bias the fitted
$\Delta E$ by $\resInitHwFitErrSitePct\,\%$ on this time grid
(Section~\ref{sec:methods}). A Gaussian packet in the left well needs no
diagonalisation; measured in the same jobs as $|L\rangle$, its splitting differs
from that of $|L\rangle$ by $\hwGaussDiffMin$ to $\hwGaussDiffMax$ percentage
points, against $\hwGaussIdealDiff$ for the noiseless circuit
(Table~\ref{tab:hw_fixed}).

At optimisation level~3 the circuit uses $\hwTwoQPerStep$ two-qubit gates
per Trotter step, the minimum $2(N-1)$, because each bond's $XX$ and $YY$
rotations are combined into a Givens rotation~\cite{kivlichan_2018,arrazola_2022};
the deepest circuit has $\hwTwoQMax$ gates. The transpiler chooses the physical
qubits per circuit: in the reference run the 28 circuits occupy
$\hwLayChainsRef$ different eight-qubit chains, and at $\hwLayMismatchRef$ of
the $\hwPoints$ time points circuits~A and~B ran on different chains (at most
$\hwLayMismatchRep$ in the repeat runs of the same day, none on \hwBackendTwo).
At four of these points A and B used the same qubits in reversed order, and only
one short-time point used a disjoint chain. Refitting the reference run on the
$\hwLayMatchedN$ points at which A and B shared a layout moves the splitting by
$\hwLayMatchedShiftPct\,\%$. A fixed initial layout would remove this variation
in future runs.

The step size and the synthesis rest on the compilation and noise-model study
in Section~S8 of the Supplementary Information. The step $\delta t=\hwDt$ keeps
the deviation of the noiseless circuit from the exact evolution
below $\hwIdealMaxDev$ at every measured time point. This
criterion is looser than the $10^{-3}$ of the projections in
Section~\ref{sec:circuit_scaling}, so the measurement does not contradict them.

\begin{table}[htbp]
\centering
\begin{tabular}{ll}
\toprule
Processors & \hwBackend, \hwBackendTwo\ (IBM Heron) \\
Runs & $\hwRepN$ on \hwBackend\ (one more excluded, Appendix~\ref{app:hw_repro}), $\hwRepNTwo$ on \hwBackendTwo \\
Sessions & 2026-09-24; 2026-10-13/14 \\
Per run & $\hwPoints$ time points $\times$ 2 circuits $\times$ $\hwShots$ shots \\
Trotter step & $\delta t=\hwDt$, $t\le\hwTmax$ \\
Two-qubit gates & $\hwTwoQPerStep$ per step, at most $\hwTwoQMax$ per circuit \\
Transpilation & optimisation level 3, fixed seed, layout chosen per circuit \\
Error suppression & dynamical decoupling (XY4), readout twirling \\
Estimator & subspace normalisation, $P_{\text{raw}}/P_{\text{subspace}}$ \\
Fit & $A\cos^2(\Delta E\,t/2)+C$, weighted by propagated shot noise \\
\bottomrule
\end{tabular}
\caption{Protocol of the device measurements. Job identifiers, calibration
snapshots, physical-qubit assignments and raw counts are part of the frozen
data (Declarations).}
\label{tab:hw_methods}
\end{table}

\begin{figure}[htbp]
\centering
\includegraphics[width=\textwidth]{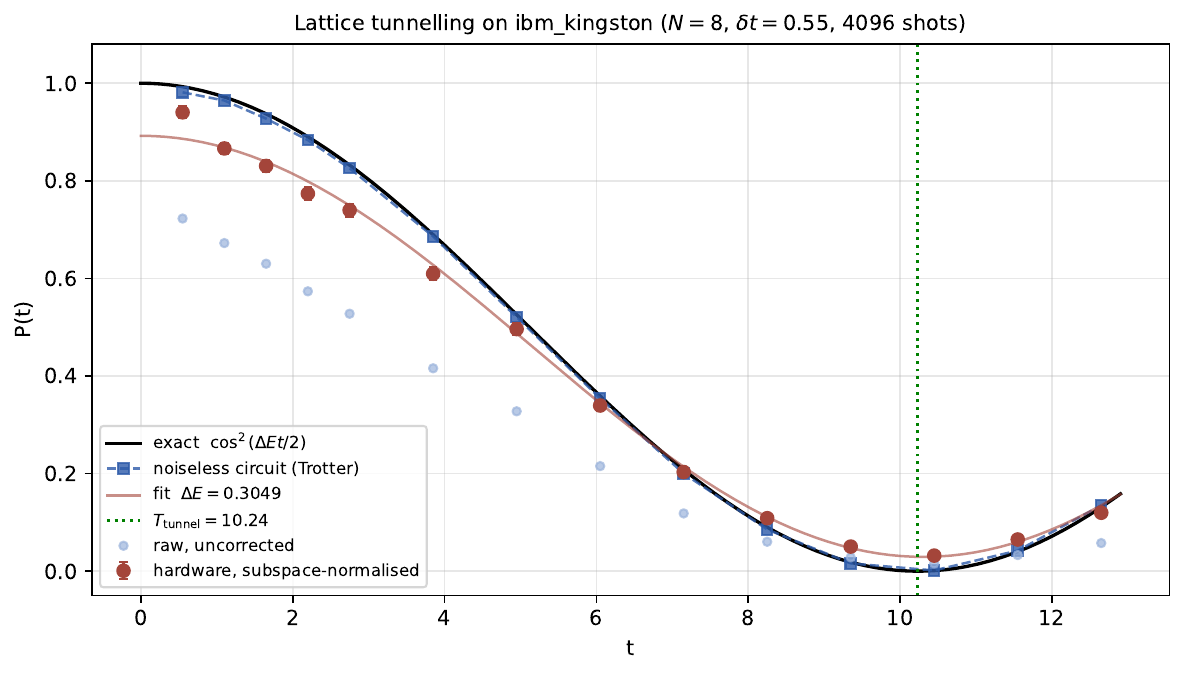}
\caption{\textbf{Tunnelling in the $N=\hwNSites$ lattice model measured on a
quantum processor} (\hwBackend). Black: exact $\cos^2(\Delta E\,t/2)$. Blue
squares: noiseless circuit at $\delta t = \hwDt$. Pale points: raw estimate.
Red: subspace-normalised estimate with shot-noise bars and the fit of
Eq.~\eqref{eq:hw_dE}. The green line marks $T_{\text{tunnel}}$. This is the
closest to the exact curve of $\hwRepN$ repetitions; their spread is given in
Eq.~\eqref{eq:hw_repro}.}
\label{fig:hw_survival}
\end{figure}

Figure~\ref{fig:hw_survival} shows one run of the measurement. Fitting
$A\cos^2(\Delta E\,t/2) + C$ to its subspace-normalised survival probability
gives
\begin{equation}\label{eq:hw_dE}
\Delta E_{\text{hardware}} = \hwDeltaE \pm \hwDeltaEErr ,
\end{equation}
against the exact $\resDeltaE$ of Method~I, a deviation of $\hwDevPct\,\%$
for this run. The quoted error is statistical only; a single run scatters by
$\pm\hwRepSd$ (Eq.~\eqref{eq:hw_repro}). Fitting the noiseless circuit and the
analytic $\cos^2$ in the same way separates the error sources: the
Trotterisation contributes $+\hwTrotterPct\,\%$ and the device
$-\hwDevicePct\,\%$. The amplitude for this run, $A = \hwAmplitude$
($\hwDevAmpKing \pm \hwDevAmpKingSd$ over the repetitions), is the residual
depolarisation that subspace normalisation does not remove. The complete
transfer is resolved, the minimum at $T_{\text{tunnel}} = \resTtunnel$ is
located, and the onset of the revival is captured.

Fitting the raw estimate without normalisation gives
$\Delta E = \hwDeltaERaw$ --- a deviation of $\hwDevPctRaw\,\%$, against
$\hwDevPct\,\%$ after normalisation --- and an amplitude of $\hwAmpRaw$
against $\hwAmplitude$. Over the $\hwRepN$ repetitions the deviation is $\hwRawAbsDevMean\,\%$
without and $\hwPostAbsDevMean\,\%$ with normalisation, because the variation
between sessions dominates both. The normalisation improves the amplitude and the depth
over which the signal survives (Appendix~\ref{app:hw_control}), not the splitting.

Circuit~B also measures the population remaining in the one-excitation sector:
\begin{equation}\label{eq:hw_leakage}
P_{\text{subspace}}(t) = \hwLeakIntercept\,e^{-t/\tau},
\qquad \tau = \hwLeakTau \pm \hwLeakTauErr ,
\end{equation}
for this run, falling from $\hwLeakIntercept$ to $\hwLeakLast$. Across the
$\hwRepN$ repetitions, the mean loss per Trotter step is
$\hwLeakKingMean \pm \hwLeakKingSd\,\%$. The intercept includes preparation
and readout error; the exponential describes loss during evolution.

ZNE restores the fitted amplitude to $\hwZneAmp$, $\hwZneAmpSigma\,\sigma$ above the
$\hwDevAmpKing \pm \hwDevAmpKingSd$ of the repeated sampler runs.
The splitting nonetheless comes out no more accurate: $\hwZneDeltaE$, a deviation of
$-\hwZneDevPct\,\%$, against $\hwSamplerDevMean \pm \hwSamplerDevSd\,\%$ for the
sampler runs (both fitted unweighted, since negative extrapolated values admit no binomial
weights), $\hwZneSigma\,\sigma$ from their mean. ZNE extrapolates each time
point independently, which restores the amplitude but not the frequency; it
also produces negative projector values and costs about three times the
processor time.

Subspace normalisation corrects leakage out of the sector, but not noise
that redistributes population within it. The inverse participation ratio
$\mathrm{IPR}=1/\sum_i p_i^2$ tracks the tunnelling at moderate depth and drifts
toward the uniform value $N$ as noise dominates, visibly from
$P_{\text{subspace}}\approx 1/3$ (Figure~\ref{fig:hw_validity}). At fixed depth
it scatters by $\hwValIprSpreadPct\,\%$ across repetitions, while over the depth
range of the figure it grows by a factor of about $\hwValIprRise$. The two runs
combined in the figure were taken $\hwValGapMin$ minutes apart, outside the
verified window of Table~\ref{tab:hw_gaps}. Across all repetitions, over the
$\hwSecNPoints$ time points, the site distribution within the sector deviates
from the noiseless one by a median total-variation distance of
$\hwSecTvdHigh$ while $P_{\text{subspace}}\ge0.5$, $\hwSecTvdMid$ between
$1/3$ and $0.5$, and $\hwSecTvdLow$ below $1/3$, a band reached only by
$\hwSecNLow$ points, all on \hwBackendTwo.

\begin{figure}[htbp]
\centering
\includegraphics[width=0.82\textwidth]{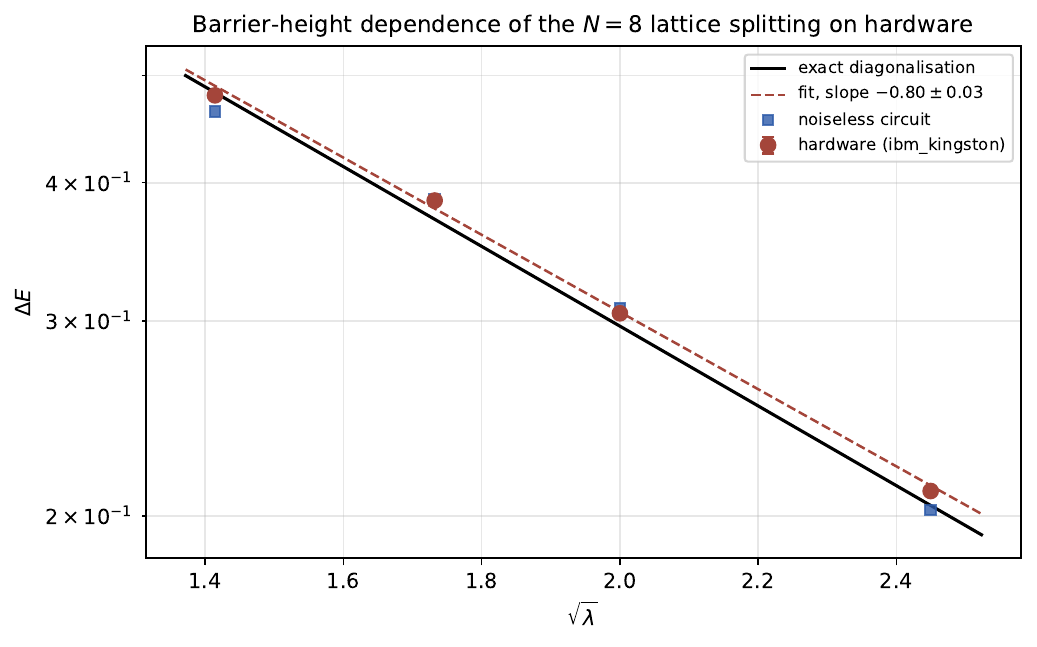}
\caption{\textbf{Barrier-height dependence measured on a processor}
(\hwBackend). Energy splitting against
$\sqrt\lambda$ on a logarithmic scale. Black: straight-line fit to exact
diagonalisation at the four $\lambda$. Blue squares: the noiseless circuit at
the $\delta t$ chosen for each $\lambda$. Red: the hardware measurement with
shot-noise bars, and the weighted fit whose slope is quoted in the text, with
its drift-inclusive error. On the $N=\hwNSites$ lattice the slope is less than
half the instanton value (Section~\ref{sec:instanton_scaling}).}
\label{fig:hw_signature}
\end{figure}

The splitting differs between \hwBackend\ and a second processor by
$\hwDevDeltaEPct\,\%$, the fitted amplitude more clearly:
$\hwDevAmpKing \pm \hwDevAmpKingSd$ over $\hwRepN$ repetitions on the first,
against $\hwAmpTwo \pm \hwAmpTwoSd$ over $\hwRepNTwo$ on \hwBackendTwo. Because runs
within a session are correlated (below), we compare session means, with
$\hwSessN$ sessions on the first processor and $\hwSessNTwo$ on the second.
A Welch test on the session means gives $t=\hwDevAmpSigmaSess$ for the
amplitude, with $\hwDevAmpDofSess$ degrees of freedom (Satterthwaite) and
$p\approx\hwDevAmpPSess$, and $t=\hwDevDeltaESigmaSess$ for the splitting, with
$\hwDevDeltaEDofSess$ degrees of freedom and $p\approx\hwDevDeltaEPSess$: the
processors differ in the damping, whereas a difference in the splitting is not
established. With so few sessions the test has little power, and the one
comparison free of session variation, alternating runs in one session, gives a
difference of $\hwJointDeltaEPct\,\%$ (Appendix~\ref{app:hw_repro}).
Treating every run as independent would give $\hwDevAmpSigma$ and
$\hwDevDeltaESigma$ standard errors and overstate both. The amplitude difference
has the expected direction: the median two-qubit error of \hwBackendTwo\ was
$\hwTwoQErrorTwo\times10^{-3}$. Repeating the protocol $\hwRepN$ times on the first
processor, on two days nearly three weeks apart, gives
\begin{equation}
\Delta E = \hwRepMean \pm \hwRepMeanSe \quad (\text{standard error, sessions as sampling unit}),
\label{eq:hw_repro}
\end{equation}
against the exact $\resDeltaE$ of Method~I. A single run scatters by
$\hwRepSd$ ($\hwRepRelPct\,\%$, standard deviation over the $\hwRepN$ runs);
$\hwRepInflation$ times the mean statistical error ($\hwRepStatPct\,\%$).
Runs taken minutes apart agree to shot noise, while the means of two sessions
differ by $\hwSessShiftPct\,\%$: the runs fall into $\hwSessN$ sessions, scatter
by $\hwSessWithinPct\,\%$ within them and by $\hwSessBetweenPct\,\%$ between
session means, so the standard error of the mean splitting is
$\hwRepMeanSePct\,\%$. This reproducibility spread, not the single-run error
bar, qualifies the measured splitting (Appendix~\ref{app:hw_repro}). A later
campaign on a fixed chain of qubits shows that this variation is not caused by
the qubit assignment alone: on identical qubits the splitting moved by
$\hwFixHourShiftPct$ percentage points within $\hwFixHourGapMin$ minutes, and a
layout chosen by the transpiler changed it by $\hwFreeVsFixPct$ points
(Table~\ref{tab:hw_fixed}). As a
control, Appendix~\ref{app:hw_control} applies the same protocol to an Ising
chain without a conserved sector (Appendix~\ref{sec:tier_f}).

Repeating the measurement across
$\lambda=\hwLamMin\ldots\hwLamMax$ reproduces the barrier-height dependence of
the $N=\hwNSites$ lattice (Figure~\ref{fig:hw_signature}). The fitted slope, $-\hwSlope\pm\hwSlopeDriftErr$ (shot noise alone would
give $\hwSlopeErr$), agrees with exact diagonalisation of the same lattice with
the same weights, $-\hwSlopeExactW$; an unweighted fit gives $-\hwSlopeExact$,
whose formal error $\hwSlopeExactErr$ measures the curvature of $\ln\Delta E$ in
$\sqrt\lambda$. Both are less than half the instanton slope $-\hwSlopeTheory$:
at this resolution the lattice sets the slope
(Section~\ref{sec:instanton_scaling}).

\section{Scope and Limitations}\label{sec:limitations}

\subsection{Scope of the Results}\label{sec:scope}

The principal physics and algorithm-validation results are classical calculations: an independent numerical confirmation of the one-loop fluctuation determinant, validation of the Trotterised circuit against the target spectrum, the step-size scaling $\delta t^{*}\propto N^{\resTierEDtStarExponentRefined}$ of the bond-wise formula and its absence for the kinetic/potential splitting, and the bound $\chi\le4$ on the bond dimension of this unary circuit. The device measurements of Section~\ref{sec:hw_campaign} resolve the tunnelling transfer at $N=\hwNSites$ with a coarser step and a separately bounded Trotter error. Hardware simulation at the paper's target accuracy is not established: the Tier~D resource estimates, projections from a simplified per-gate error model rather than executed circuits (Section~\ref{sec:circuit_scaling}), predict that a full tunnelling period at the required step size is out of reach, and the device measurements do not overturn that projection. Because $|L\rangle$ is prepared from the exact eigenvectors, the device measures the phase accumulated by a prepared two-level superposition; a Gaussian start without classical diagonalisation gives the same splitting to within $\hwGaussDiffAbsMax$ percentage points (Table~\ref{tab:hw_fixed}).

This work studies a single particle in a quartic double well. The unary encoding confines the dynamics to the one-excitation sector, so the model remains classically tractable; the results validate the algorithm but do not demonstrate quantum advantage. The bond-dimension bound and the $N^{\resTierECostExponentRefined}$ cost apply to this encoding and the bond-wise Trotter synthesis, not to alternative encodings or splittings (Sections~\ref{sec:tn_chi} and~\ref{sec:tn_dt_scaling}). The device measurements are made at $N=\hwNSites$, where the lattice splitting lies well below the continuum value and its $\lambda$-dependence is not the instanton one. Tier~F is a control on a different Hamiltonian and does not extend the double-well physics claims.

Validation is limited to the lattice sizes reached by the statevector and tensor-network calculations; extrapolations beyond them remain projections. The continuum uncertainty reflects fit-window and model choices (Appendix~\ref{sec:extrapolation_robustness}), and the semiclassical comparison includes the one-loop determinant but neither higher-loop terms nor interactions beyond the dilute-gas approximation (Appendix~\ref{app:determinant_check}). The numerical uncertainty budget of the pipeline (Appendix~\ref{app:budget}) is dominated by method-limited contributions: the determinant extrapolation, the continuum fit and the Trotter step. Hardware reproducibility is not part of that budget; its spread, $\hwRepRelPct\,\%$ in $\Delta E$ (Eq.~\eqref{eq:hw_repro}), is reported with the device measurement in Section~\ref{sec:hw_campaign} and Appendix~\ref{app:hw_repro}.

\subsection{Benchmark Protocol}\label{sec:benchmark_protocol}

The benchmark is meant to be run elsewhere, so we state it as a protocol. A
platform prepares $|L\rangle$ on the $N$-site lattice (its two eigenvectors
are an $N\times N$ tridiagonal problem at any $N$), applies the
odd/even circuit of Section~\ref{sec:hw_campaign} at a stated step $\delta t$,
and measures the two circuits A and B at each time point. Three numbers
characterise the run, each with an exact or reproducible reference:
\begin{enumerate}
\item the deviation of the fitted $\Delta E$ from the exact lattice value of
Method~I, which the pipeline (Appendix~\ref{app:reproducibility}) returns for any $N$ and $\lambda$, reported
against the number of two-qubit gates;
\item the surviving-signal scale $n_0$ of the subspace-normalised estimator
(Eq.~\eqref{eq:hw_F}), and its ratio to the unnormalised one;
\item the reproducibility: repetitions in at least two sessions separated by
more than the verified window of Appendix~\ref{app:hw_repro}, with the standard
error formed over session means.
\end{enumerate}
For comparable results we recommend the settings of the present measurement:
a step $\delta t$ for which the noiseless circuit stays within $\hwScanTol$ of
the exact evolution, time points out to the onset of the revival, the fit
$A\cos^2(\Delta E\,t/2)+C$ weighted by shot noise, and $n_0$ as defined in
Eq.~\eqref{eq:hw_F}, together with the physical qubits used and their
calibration. Cost should be reported both as two-qubit gate count and as
two-qubit depth, since platforms with all-to-all connectivity need no routing.
At $N=\hwNSites$ on \hwBackend\ these are a mean deviation within
$\hwRepMeanSePct\,\%$ (Eq.~\eqref{eq:hw_repro}) at up to 354 two-qubit gates,
$n_0=\hwIsNzeroPostMean\pm\hwIsNzeroPostSd$ with a gain of
$\hwIsRatioMean\pm\hwIsRatioSd$, and a session-to-session spread of
$\hwSessBetweenPct\,\%$. The lattice values are not the continuum instanton
splitting: at $N=\hwNSites$ they lie $\resConvRelErrNSites\,\%$ below it (Table~\ref{tab:convergence}, Section~\ref{sec:n8_scope}),
and the benchmark approaches the instanton problem only as $N$ grows past the
resolved grids.

\section{Conclusion and Outlook}\label{sec:conclusion}

We mapped the double-well Schrödinger equation onto a qubit lattice by a
unary encoding and validated the Trotterised circuit against exact
diagonalisation. Every classical number in the paper can be checked against an
exact answer; the uncertainties are collected in Appendix~\ref{app:budget}.

The exact continuum splitting exceeds the zero-mode-only formula
$\Delta E_0$ by a factor $\resRatioInfSc$. This gap is not a quantum effect
beyond the semiclassical description: $\Delta E_0$ omits the fluctuation
determinant, and restoring it, as computed numerically here, multiplies
$\Delta E_0$ by $\sqrt{12}$. The resulting one-loop value $\Delta E_1$ lies
\resOneLoopOverExactPct\,\% above the exact splitting at the benchmark
parameters, and the difference shrinks as the barrier grows, as expected for
higher-order corrections (Section~\ref{sec:determinant_result} and
Appendix~\ref{app:determinant_check}).

The Trotterised circuit reproduces the spectrum to \resTierEWorstRel\ up to
$N=48$. Its bond dimension never exceeds four, independently of $N$
(Lemma~\ref{prop:chi4}), and its non-stabilizerness grows only
logarithmically with $N$ (Appendix~\ref{sec:tn_magic}). For the bond-wise
synthesis the admissible step falls roughly as $N^{-2}$, robustly against the
accuracy criterion, so
one tunnelling period costs about $N^{\resTierECostExponentRefined}$ two-qubit
gates; a kinetic/potential splitting keeps the step independent of $N$ and
lowers the exponent to two. Either way the simulation is limited by circuit
depth, not by qubit number.

A full tunnelling period at the paper's accuracy criterion is out of reach on
current hardware at every $N$ considered. At a coarser, separately bounded step
we resolve the complete transfer at $N=\hwNSites$ on two superconducting
processors, recover the lattice splitting to a few percent and reproduce its
slope against $\sqrt\lambda$ ($-\hwSlope\pm\hwSlopeDriftErr$ against
$-\hwSlopeExactW$). Two lessons extend beyond this model. Normalising by the
measured population of the conserved sector extends the usable depth by a factor
of about $\hwIsRatioRound$ ($\hwIsRatioMean\pm\hwIsRatioSd$), but does not make
the splitting more accurate. The dominant uncertainty is the variation between runs taken hours or more
apart: runs minutes apart agree to shot noise, sessions differ by up to
$\hwSessShiftPct\,\%$, and a single-run error bar understates the uncertainty by
a factor of about $\hwRepInflation$. Temporal drift and the qubit assignment
contribute with comparable size.

The unary double well is a benchmark for correctness, not for cost; questions
of cost scaling require models with entanglement growth, such as the
false-vacuum Ising chain of Tier~F.

\subsection{Outlook}\label{sec:future_outlook}

Extended-precision diagonalisation at larger action would test the
three-loop coefficient~\cite{escobar_ruiz_2015} directly.

Two routes address the depth bottleneck: keeping the kinetic term whole
(Section~\ref{sec:tn_dt_scaling}) and post-Trotter methods such as
qubitization~\cite{low_chuang_2019}. On the readout side, the short-time decay
of $P_L$ or phase estimation of the relative phase between $|\psi_0\rangle$ and
$|\psi_1\rangle$ would avoid resolving a full period. Beyond the
single-excitation sector the initial state must be prepared without
eigenvectors; a smooth packet in one well keeps most of its weight in the
doublet (Section~\ref{sec:methods}).

On the hardware side, mid-circuit measurement of the excitation number would turn subspace
normalisation into genuine error detection, and the Givens-rotation synthesis
is the natural starting point for short-time measurements at $N=12$--$16$.

The methodology can be carried into models where the classical tiers stop being
cheap. Two particles in a
double well, or a Bose--Hubbard dimer, add interactions while remaining exactly
solvable for validation. In $\phi^4$ theory in $1+1$ dimensions, the two vacua
of the $\mathbb{Z}_2$-broken phase are split in a finite volume $L$ by tunnelling
through kink configurations, with $\Delta E\propto e^{-M_{\text{kink}}L}$ and a
one-loop determinant prefactor~\cite{rychkov_vitale_2016} --- the field-theoretic
counterpart of the instanton splitting and the determinant studied here, in a
setting where entanglement does grow. Tier~F already shows that the same tools resolve bond-dimension growth in the
false-vacuum Ising chain.

\backmatter

\bmhead{Supplementary information}
The Supplementary Information contains, in Sections~S1--S6, the step-by-step
semiclassical derivations behind Section~\ref{sec:theory} and
Appendix~\ref{app:semiclassics}: Euclidean path integral, kink solution,
zero-mode Jacobian, fluctuation determinant, dilute-gas summation and WKB
comparison. Section~S7 contains the unary subspace and the construction of the
localised states (Section~\ref{sec:lattice}), and Section~S8 the compilation,
coherence and calibrated-noise study that preceded the hardware campaign
(Section~\ref{sec:outlook}).

\section*{Declarations}

\begin{itemize}
\item \textbf{Funding:} The work of G.\,H. is supported by the Ministry of
Economic Affairs, Labour and Tourism Baden-W\"urttemberg in the frame of the
Competence Center Quantum Computing Baden-W\"urttemberg (Project No.\
KQCBW25) and by the German Federal Ministry of Research, Technology and Space
within the funding program `Application-oriented quantum computing' under
Contract No.\ 13N17159.
\item \textbf{Quantum computing resources:} We acknowledge the use of IBM
Quantum services within the free plan for this work.
\item \textbf{Competing interests:} M.\,M. is employed by infoteam Software AG. G.\,H. declares that the research was conducted in the absence of any commercial or financial relationships that could be construed as a potential conflict of interest.
\item \textbf{Ethics approval and consent to participate:} Not applicable.
\item \textbf{Consent for publication:} Not applicable.
\item \textbf{Data availability:} The frozen device data --- raw counts, job
identifiers, calibration snapshots and physical-qubit assignments --- and all
data underlying the figures and tables are deposited in a self-contained Zenodo
replication archive, which is made available to the editor and referees for
review; its DOI will be added on acceptance.
\item \textbf{Materials availability:} Not applicable.
\item \textbf{Code availability:} The complete simulation and analysis pipeline,
together with an automated test suite that checks the results against stored
reference values, is deposited in the same Zenodo replication archive.
\item \textbf{Author contribution:} G.\,H.: conceptualization, methodology,
software, formal analysis, writing -- original draft, writing -- review \&
editing. M.\,M.: software, investigation, writing -- review \& editing.
\end{itemize}

\begin{appendices}
\counterwithin*{figure}{section}
\counterwithin*{table}{section}
\counterwithin*{equation}{section}
\renewcommand{\theHfigure}{\Alph{section}.\arabic{figure}}
\renewcommand{\theHtable}{\Alph{section}.\arabic{table}}
\renewcommand{\theHequation}{\Alph{section}.\arabic{equation}}

\section{Methods: Software, Reproducibility and Use of AI Tools}\label{app:reproducibility}

Every figure, table, and macro-bound numerical result is produced by one
pipeline run in Python, using Qiskit, Qiskit Aer and quimb; the device measurements are frozen
as raw counts and re-analysed by the same code. The
deposited test suite of 205 checks covers closed-form physics benchmarks, the
pipeline results against stored reference values, the lemma, and the analysis
of the frozen device data. The tensor-network tests are marked as slow and can be deselected.

\noindent\textbf{Use of AI tools.} During the preparation of this work the
authors used Claude (Anthropic) and Antigravity (Google) in order to assist with analysis code, consistency checks and language
editing.
The authors reviewed and edited the output and take full responsibility for
the content of the publication. 

\section{Numerical Uncertainty Budget}\label{app:budget}

Table~\ref{tab:budget} collects the relative uncertainty of each step of the
numerical pipeline. The device measurements are excluded; their reproducibility
is discussed in Appendix~\ref{app:hw_repro}.

\begin{table}[ht]
\centering
\begin{tabular}{|l|c|l|}
\hline
contribution & relative & character \\
\hline
MPS truncation at $\chi=4$ & $0$ (exact) & exact \\
Magic estimator (replica vs exact) & $5.1\times10^{-15}$ & numerical \\
Eigensolver residual & $3.8\times10^{-13}$ & numerical \\
Box size at $L=3$ & $9.9\times10^{-13}$ & numerical \\
Tier F TEBD vs dense diagonalisation & $4.1\times10^{-6}$ & numerical \\
Continuum extrapolation of $\Delta E$ & $5.4\times10^{-5}$ & method-limited \\
Trotter error after Richardson in $\delta t$ & $6.6\times10^{-5}$ & method-limited \\
One-loop determinant, two routes & $5.1\times10^{-4}$ & method-limited \\
\hline
\end{tabular}
\caption{Relative uncertainty of every step that feeds a number quoted in this paper, ordered by size. Each entry is measured, not estimated, and the table is assembled from those measurements rather than recomputed. The three method-limited entries span \resDEInfUncertaintyRel\ to \resBudgetDominant\ and are each limited by a choice rather than by precision: the fit window and form for the continuum extrapolation (Appendix~\ref{sec:extrapolation_robustness}), the affordable step size for the Trotter error, and the range of $S_0/\hbar$ available to the $1/S_0$ extrapolation for the determinant. Every numerical entry sits at least an order of magnitude lower, and the MPS truncation is identically zero rather than small, because the Schmidt rank never exceeds \resRankMaxIntermediate\ (Lemma~\ref{prop:chi4}). $\Delta E_{sc}$ is not in this table: the leading-order convention is a definition, not an uncertainty.}
\label{tab:budget}
\end{table}

\section{Semiclassical Derivations}\label{app:semiclassics}

This appendix checks the fluctuation determinant behind $\Delta E_1$ (Eq.~\eqref{eq:one_loop}) numerically and compares with WKB. The derivations of $\Delta E_0$ and $\Delta E_1$, following Coleman~\cite{coleman_aspects} and Zinn-Justin~\cite{zinn_justin_QFT}, are given in Sections~S1--S6 of the Supplementary Information.

\subsection{Numerical Check of the Fluctuation Determinant}\label{app:determinant_check}

\begin{figure}[htbp]
\centering
\includegraphics[width=0.9\textwidth]{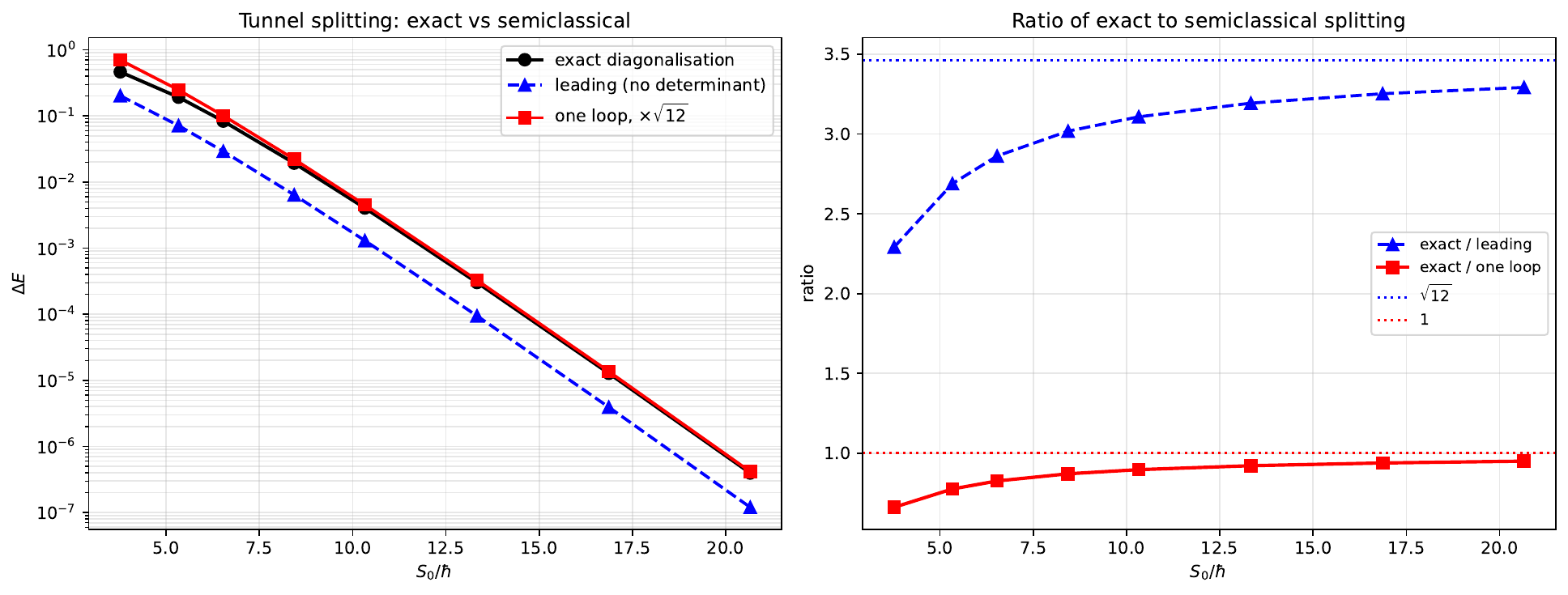}
\caption{Fluctuation-determinant check over six orders of magnitude in the
splitting. \textbf{Left:} Exact diagonalisation (black), $\Delta E_0$ (blue
dashed) and $\Delta E_1=\sqrt{12}\,\Delta E_0$ (red).
\textbf{Right:} $\Delta E_{\text{exact}}/\Delta E_{\text{leading}}$ approaches
$\sqrt{12}=\resDetOneLoopPrefactor$, while
$\Delta E_{\text{exact}}/\Delta E_1$ approaches unity from below.}
\label{fig:determinant}
\end{figure}

For $V=\lambda(x^2-a^2)^2$, the instanton is
$x_{\text{inst}}(\tau)=a\tanh(\omega\tau/2)$. With $u=\omega\tau$ and
$s=u/2$, its fluctuation operator is the P\"oschl--Teller operator
\begin{equation}
\hat M = \frac{\omega^2}{4}\left[-\partial_s^2 + 4 - 6\,\mathrm{sech}^2 s\right].
\end{equation}
with a translation zero mode, a bound state at $3\omega^2/4$ and a continuum
above $\omega^2$. We evaluate the determinant in two independent ways. In the
Gel'fand--Yaglom calculation, integrating
$\psi''=\omega^{-2}V''(x_{\text{inst}})\psi$ with
$\psi(-U/2)=0$, $\psi'(-U/2)=1$, and removing the zero mode gives
\begin{equation}
\frac{\det{}'(\hat M)}{\det(\hat M_0)} = \resDetOneLoopRatio
\qquad (\text{exact value } 1/12,\ \text{in units of } \omega^{-2}),
\label{eq:gy_result}
\end{equation}
recovered to $2.4\times10^{-5}$ relative. Independently, fitting
$\Delta E_{\text{exact}}/\Delta E_0$ over $\lambda=4\ldots120$ and
extrapolating to large action over $S_0/\hbar\geq\resDetFitSZeroMin$ gives
$\resDetRatioExactOverLeading$. The two determinations agree to $0.05\%$ (Figure~\ref{fig:determinant}).
The exact splittings are extrapolated in $1/N^2$ from grids of 1600, 3200
and 6400 points. At the largest action ($\lambda=120$,
$\Delta E\approx4\times10^{-7}$) the grids agree to $2\times10^{-3}$, close to
the double-precision floor of the eigensolver, so the comparison there is
meaningful to about $10^{-3}$.

At the standard parameters $\Delta E_\infty/\Delta E_1 = \resRatioInfOneLoop$.
Across $S_0/\hbar=\resSZero\ldots20.7$, the measured deficit
$(1-\Delta E_{\text{exact}}/\Delta E_1)S_0/\hbar$ falls from
$1.28$ to $1.02$. The known two-loop expansion
$\delta E=\Delta E_1\sum_{n\geq0}B_n S_0^{-n}$ has
$B_0=1$ and $B_1=-71/72$ for this potential and convention
\cite{wohler_shuryak_1994,escobar_ruiz_2015}. It predicts a ratio of $0.952$
at $S_0/\hbar=20.7$, matching the measured $0.95$ to $0.2\%$; at the standard
action it predicts $0.739$, the larger discrepancy being due to higher orders.

The low-lying-state WKB estimate~\cite{garg_2000}, with $E_0=\hbar\omega/2$ in
the barrier exponent, the integral taken between the inner turning points and
the prefactor $\hbar\omega/\pi$, gives $\Delta E_{\text{WKB}}=\resDEWkbCorr$,
within $0.3\%$ of $\Delta E_1=\resDEScOneLoop$, as Garg's analysis leads one to
expect.

\FloatBarrier
\section{Robustness of the Continuum Extrapolation}\label{app:extra}\label{sec:extrapolation_robustness}

\begin{table}[ht]
\centering
\begin{tabular}{|l|c|c|c|}
\hline
\multicolumn{4}{|l|}{\textbf{(a) Fit window}, linear in $1/N^2$} \\
\hline
window & $\Delta E_\infty$ & $c_2$ & deviation \\
\hline
$12\ldots48$ & 0.455476 & -8.94 & $-1.503\%$ \\
$16\ldots64$ & 0.463463 & -14.67 & $+0.224\%$ \\
$24\ldots96$ & 0.462648 & -13.61 & $+0.048\%$ \\
$32\ldots128$ & 0.462515 & -13.28 & $+0.019\%$ \\
$48\ldots192$ & 0.462453 & -12.96 & $+0.005\%$ \\
$64\ldots256$ & 0.462438 & -12.82 & $+0.002\%$ \\
$96\ldots384$ & 0.462431 & -12.67 & $+0.001\%$ \\
\hline
\multicolumn{4}{|l|}{\textbf{(b) Fit form}, same windows} \\
\hline
window & one term & two terms & $c_4$ \\
\hline
$12\ldots48$ & 0.455476 ($-1.503\%$) & 0.469001 ($+1.421\%$) & 1861 \\
$24\ldots96$ & 0.462648 ($+0.048\%$) & 0.462462 ($+0.007\%$) & -409 \\
$48\ldots192$ & 0.462453 ($+0.005\%$) & 0.462433 ($+0.001\%$) & -713 \\
\hline
\multicolumn{4}{|l|}{\textbf{(c) Uncertainty budget} (relative, on the configured window)} \\
\hline
\multicolumn{3}{|l|}{deviation from the fine-grid reference} & $5.4\times10^{-5}$ \\
\multicolumn{3}{|l|}{fit window} & $4.7\times10^{-5}$ \\
\multicolumn{3}{|l|}{fit form (one term vs two)} & $4.4\times10^{-5}$ \\
\multicolumn{3}{|l|}{leave-one-out} & $4.4\times10^{-5}$ \\
\multicolumn{3}{|l|}{box size} & $9.9\times10^{-13}$ \\
\multicolumn{3}{|l|}{eigensolver residual} & $3.8\times10^{-13}$ \\
\hline
\end{tabular}
\caption{Robustness of the continuum extrapolation. \textbf{(a)} Sliding the fit window: the limit is approached monotonically once the wells are resolved, and the configured window $\resRichardsonWindow$ reproduces the $\Delta x = \resRichardsonReferenceDx$ reference to \resRichardsonVsRefRel. \textbf{(b)} Adding an $N^{-4}$ term. It helps only where the window can constrain it: on $12\ldots48$ the error merely changes sign without changing size, and $c_4$ comes out positive where the resolved windows agree on $\approx-700$ -- there the second coefficient absorbs the unmodelled $N^{-6}$ remainder instead of measuring $N^{-4}$. \textbf{(c)} The resulting budget. The three model-choice entries agree within a factor of two and sit nine orders of magnitude above the arithmetic, so the limit is set by how the extrapolation is done and not by how well it is computed. No statistical term appears: $\Delta E(N)$ is exact diagonalisation, deterministic to $10^{-13}$, so its fit residuals are the systematic $O(N^{-4})$ term and not noise.}
\label{tab:extrapolation_robustness}
\end{table}

Table~\ref{tab:extrapolation_robustness} varies the fit window, the fit form and the points in turn. Sliding the window (a) shows a monotone approach once the wells are resolved: $+0.22\%$ at $16\ldots64$, $+0.05\%$ at $24\ldots96$, $+0.005\%$ at the configured $\resRichardsonWindow$ and $+0.001\%$ at $96\ldots384$. An added $N^{-4}$ term (b) improves the fine windows by a further factor of five, but on $12\ldots48$ it only flips the sign of the error, from $-1.5\%$ to $+1.4\%$, with $c_4$ of the wrong sign: there the second coefficient absorbs the unmodelled higher orders. Dropping any single point from the configured window moves the limit by at most \resLooSpreadRel. The largest measure is the \resRichardsonVsRefRel\ difference from the value computed directly at $\Delta x = \resRichardsonReferenceDx$, which Eq.~\eqref{eq:dEinf_quoted} quotes.

The reference value $\resDEContinuumRef$, a limit in $\Delta x$ at fixed box, lies inside this interval. We quote the fit because it is what the resolution study of Table~\ref{tab:convergence} and a device sampling those lattices yield; the ratio to $\Delta E_0$ is $\resRatioInfSc$ either way. No statistical uncertainty is quoted: $\Delta E(N)$ is deterministic, and the fit residuals are the systematic higher-order terms, not noise.

\section{Magic and Bond Dimension}\label{sec:tn_magic}

Lemma~\ref{prop:chi4} bounds the entanglement of the simulated states, not their non-stabilizerness (``magic''), which governs the cost of stabilizer-based simulation via the stabilizer rank~\cite{bravyi_gosset_2016}.

\begin{figure}[htbp]
\centering
\includegraphics[width=\textwidth]{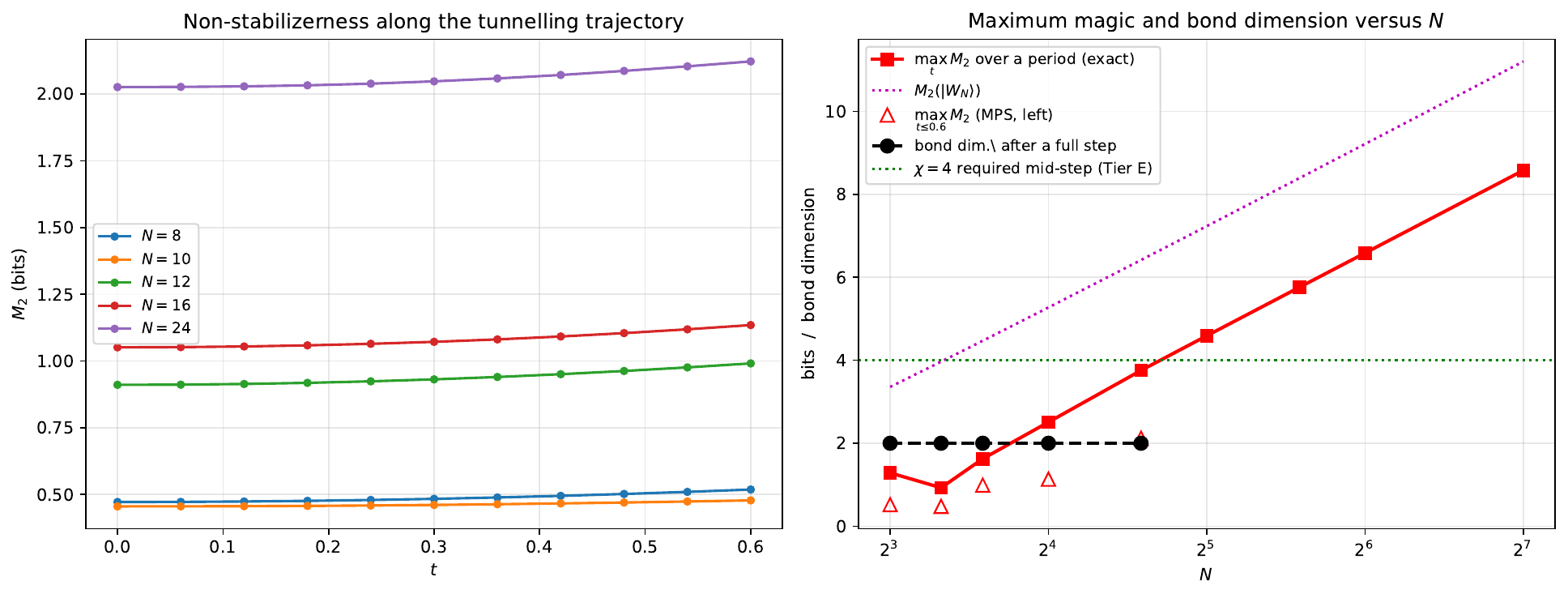}
\caption{\textbf{Left:} $M_2(t)$ from the MPS replica estimate over the start of the tunnelling trajectory for several $N$. \textbf{Right:} $\max_t M_2$ over a full period from Eq.~\eqref{eq:m2_unary_exact} on the exact trajectory (squares), $M_2$ of the $W$ state (dotted), the short-window MPS maximum (open triangles) and the bond dimension after a complete step.}
\label{fig:magic}
\end{figure}

We quantify magic by the stabilizer $2$-R\'enyi entropy $M_2$~\cite{leone_2022}, evaluated on the MPS by a four-replica transfer contraction~\cite{haug_piroli_2023} and cross-checked against both the exact $4^N$ Pauli spectrum for $N\leq12$ and the exact $\mathcal{O}(N)$ single-excitation closed form:
\begin{equation}\label{eq:m2_unary_exact}
M_2\!\left(\sum_{k=1}^N c_k |e_k\rangle\right) = -\log_2\!\left[6\left(\sum_{k=1}^N |c_k|^4\right)^2 - 6\sum_{k=1}^N |c_k|^8 + \left|\sum_{k=1}^N c_k^4\right|^2\right],
\end{equation}
obtained by summing the diagonal $Z$-strings and the weight-two $X/Y$ strings analytically (agreement to $10^{-15}$). For a uniform $W$ state it gives $M_2(|W_N\rangle) = 3\log_2 N - \log_2(7N-6) \sim 2\log_2 N$. Since the state lies in the one-excitation sector after every complete Trotter step, Eq.~\eqref{eq:m2_unary_exact} gives $M_2$ along a whole period at any $N$; the replica contraction is affordable only over its start (Table~\ref{tab:magic}, Figure~\ref{fig:magic}). $M_2$ peaks near $T/2$, and its maximum grows as $\resMagicPeriodLogSlope\,\log_2 N$ for $N\geq16$, logarithmically like a $W$ state. The short MPS window alone would suggest a power law.

\begin{table}[ht]
\centering
\begin{tabular}{|c|c|c|c|c|}
\hline
$N$ & $\max_t M_2$, period & $t_{\max}/T$ & $M_2(|W_N\rangle)$ & $\max_{t\leq0.6} M_2$, MPS \\
\hline
8 & 1.286 & 0.61 & 3.356 & 0.518 \\
10 & 0.927 & 0.73 & 3.966 & 0.478 \\
12 & 1.625 & 0.65 & 4.469 & 0.991 \\
16 & 2.508 & 0.50 & 5.272 & 1.134 \\
24 & 3.755 & 0.50 & 6.415 & 2.122 \\
32 & 4.590 & 0.50 & 7.232 & --- \\
48 & 5.758 & 0.50 & 8.389 & --- \\
64 & 6.585 & 0.50 & 9.212 & --- \\
128 & 8.578 & 0.50 & 11.202 & --- \\
\hline
\end{tabular}
\caption{Non-stabilizerness $M_2$ (bits) along the tunnelling evolution. Over a full period $T=\pi\hbar/\Delta E$ the maximum is reached near $T/2$, where the state is spread over both wells, and grows as 2.02$\,\log_2 N$ for $N\geq16$, the logarithmic growth of a $W$ state; evaluated with Eq.~\eqref{eq:m2_unary_exact} on the exact trajectory. The last column is the four-replica MPS estimate ($\chi=4$, $\delta t=0.01$, 60 steps), which covers only the start of the period and is close to $M_2$ of the initial state. After every complete Trotter step the bond dimension is $2$; the $\chi=4$ of Section~\ref{sec:tn_chi} is required by the intermediate states.}
\label{tab:magic}
\end{table}

The growth of magic does not make the MPS simulation harder: the bond
dimension stays at most four, so each Trotter step costs resources linear in
$N$. Stabilizer-based methods, whose cost depends on magic, may become less
efficient as $N$ grows.

\section{Tier F: False-Vacuum Decay in the Ising Chain}\label{sec:tier_f}

The bounded bond dimension of the unary double well is a property of the model, not of our methods. To show this we apply the same tools to an Ising chain with a transverse and a longitudinal field,
\begin{equation}
H = -J\sum_i Z_i Z_{i+1} \;-\; g\sum_i X_i \;-\; h\sum_i Z_i .
\end{equation}
For $h=0$ the two ferromagnetic vacua are degenerate, split only by an amount exponentially small in $N$. For $h>0$ the state polarised against the field is a false vacuum that decays by nucleation of true-vacuum bubbles, the lattice analogue of Coleman's bounce~\cite{rutkevich_1999,lagnese_2021,milsted_2022}. Entanglement after such a quench grows with time~\cite{calabrese_cardy_2005}. We evolve by TEBD, validated against dense diagonalisation to \resIsingValidationErr\ (Table~\ref{tab:ising_bond}, Figure~\ref{fig:ising}). The same quench at $N=\hwIsN$ serves as the hardware control of Appendix~\ref{app:hw_control} (Figure~\ref{fig:hw_ising}).

\begin{table}[ht]
\centering
\begin{tabular}{|c|c|c|c|c|}
\hline
$N$ & bond dim.\ at $T$ & $2^{N/2}$ (exact max) & $S_{vN}$ (bits) & $\log_2(\text{rank})$ \\
\hline
10 & 32 & 32 & 1.099 & 5.00 \\
12 & 61 & 64 & 1.287 & 5.93 \\
16 & 106 & 256 & 1.262 & 6.73 \\
\hline
\multicolumn{2}{|l|}{unary double well (Tier E)} & --- & --- & $\chi = 4$, all $N$, all $t$ \\
\hline
\end{tabular}
\caption{Bond dimension reached after a quench from the polarised state at $h=0$, the reference of the $h$ scan in Figure~\ref{fig:ising}, where no vacuum is false ($J=1$, $g=0.5$, $T=8$, $\delta t=0.05$, cutoff $10^{-12}$, cap $\chi=256$). The cap is not reached at the largest $N$, so these are the values the physics demands, not the values the truncation permits.}
\label{tab:ising_bond}
\end{table}

\begin{figure}[htbp]
\centering
\includegraphics[width=\textwidth]{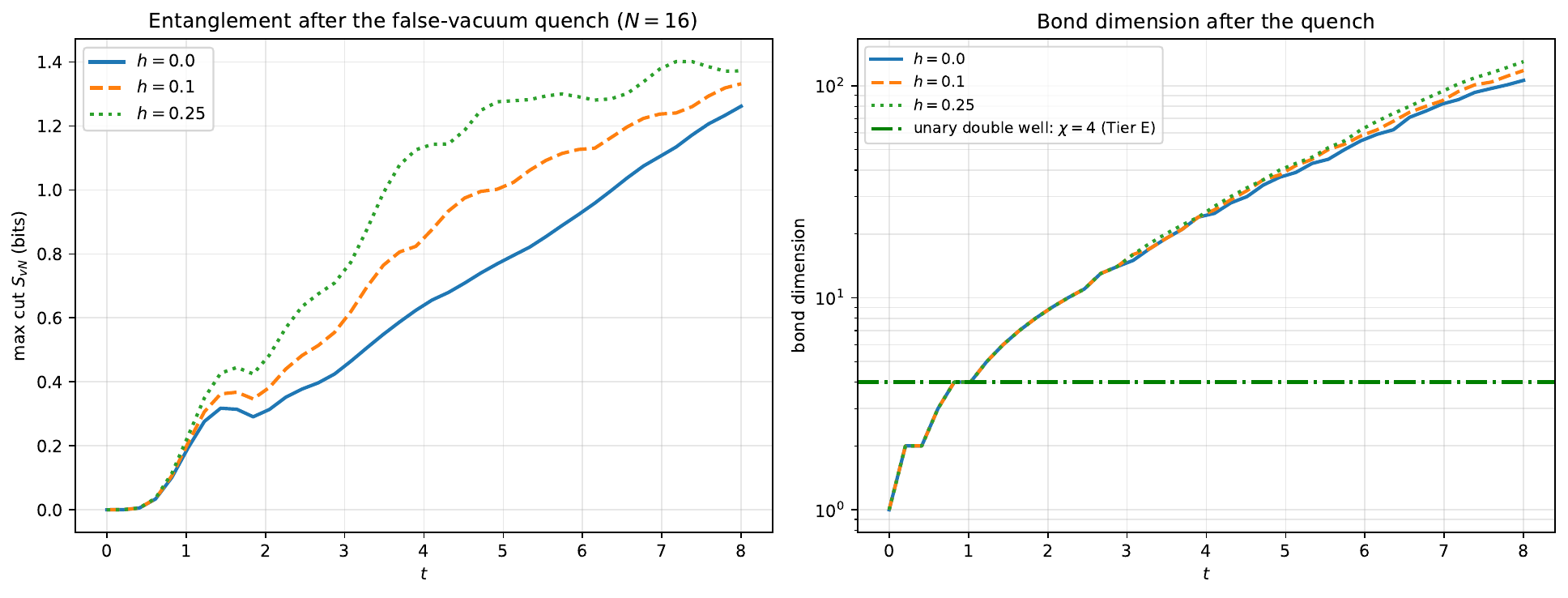}
\caption{\textbf{Left:} entanglement entropy after the false-vacuum quench at $N=16$ for three longitudinal fields. \textbf{Right:} the bond dimension over time, against the constant $\chi=4$ of the unary double well.}
\label{fig:ising}
\end{figure}

The entropy does not bound the rank: at $N=16$ the entanglement entropy is $1.26$ bits, while the rank is $106$, because a spectrum with one dominant value and a long tail has low entropy and high rank. The dependence on $h$ cannot be read as a confinement result at these sizes, since $h$ changes both the string tension and the injected energy density, and at $h=0.1$ the meson size of about 20 sites exceeds the chain.

The contrast is clear: with entanglement growth the bond dimension rises from $1$ to over $100$ within a few tunnelling times and increases with $N$, whereas in the unary double well it stays at or below four.

\section{Reproducibility of the Device Measurements}\label{app:hw_repro}

This appendix gives the repeat analysis behind Section~\ref{sec:hw_campaign}
($\hwRepN$ runs on the first processor, $\hwRepNTwo$ on the second), the
mitigation ablation and the timing of the comparisons.

The range across the $\hwRepN$ runs on the first processor,
$\hwRepRangePct\,\%$, exceeds the $\hwDevDeltaEPct\,\%$ that separates the two
processors on average. Run in alternation within one session, they differ in the
splitting by $\hwJointDeltaEPct\,\%$, but the first processor sat at the low end
of its own range that morning.

\begin{figure}[htbp]
\centering
\includegraphics[width=\textwidth]{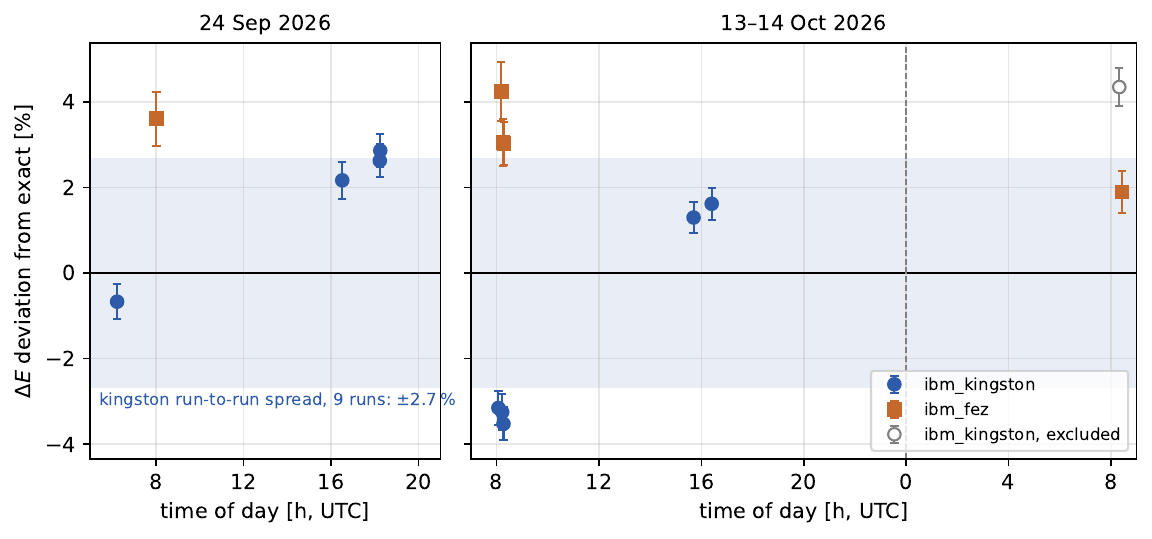}
\caption{\textbf{Drift between sessions} (\hwBackend\ and \hwBackendTwo). Deviation
of the normalised $\Delta E$ from the exact value for every run of the
identical protocol, against the time the job began executing, with shot-noise bars.
The shaded band is the run-to-run spread, $\pm\hwRepRelPct\,\%$, over the
$\hwRepN$ runs on \hwBackend. Open circle: the excluded run.}
\label{fig:hw_drift}
\end{figure}

The scatter has structure (Figure~\ref{fig:hw_drift}). Three runs taken
within $\hwVerifiedSpan$ minutes of one another spread by $\hwSessRelPct\,\%$
against an expected $\hwSessStatPct\,\%$, with $\chi^2/\text{dof} =
\hwSessChi$; three runs nearly three weeks later, within minutes of one another, give
$\chi^2/\text{dof} = \hwSessTwoChi$, both consistent with shot noise. The
means of the two sessions differ by $\hwSessShiftPct\,\%$, and a run about ten
hours before the first sits $\hwDriftPct\,\%$ from it. The published median
calibration does not reveal this: on the first day the snapshots of all four
runs are identical. With the same seed, however, the transpiler placed the
reference run and the three later runs of that day on different physical qubits,
which indicates changed per-qubit calibration data. Differences between sessions
can therefore combine temporal drift with changes of the qubit assignment.

A repeat campaign in November 2026 separates the two (Table~\ref{tab:hw_fixed}).
It runs the protocol on the fixed chain of the reference run, logs the
calibration of every qubit and coupler used, and measures $|L\rangle$ and the
Gaussian start in the same job. Two jobs $\hwFixPairGapMin$ minutes apart agree
to $\hwFixPairDiffPct$ percentage points. On the same qubits the $\hwFixN$
fixed-chain jobs span $+\hwFixMinPct$ to $+\hwFixMaxPct\,\%$, against
$\hwDevPct\,\%$ for the reference run on this chain in September, and two of them
$\hwFixHourGapMin$ minutes apart differ by $\hwFixHourShiftPct$ points. A job
with the layout chosen by the transpiler, $\hwFreeGapMin$ minutes after a
fixed-chain job, differs from it by $\hwFreeVsFixPct$ points. Temporal drift and
the qubit assignment thus contribute with comparable size, and drift can appear
within an hour. One coupler of the fixed chain was two to three times worse than
the device median in these jobs.

\begin{table}[htbp]
\centering
\caption{Repeat campaign of November 2026 on \hwBackend. Deviation of the fitted $\Delta E$ from the exact lattice value for the start states $|L\rangle$ and the Gaussian packet, measured in the same job. All jobs but the last use the fixed chain of the reference run.}
\label{tab:hw_fixed}
\begin{tabular}{llrrr}
\hline
Executed (UTC) & Layout & $|L\rangle$ [\%] & Gaussian [\%] & difference \\
\hline
2026-11-02 10:24 & fixed & $+8.66$ & $+8.13$ & $-0.53$ \\
2026-11-02 10:26 & fixed & $+8.91$ & $+8.01$ & $-0.90$ \\
2026-11-04 13:53 & fixed & $+6.40$ & $+4.85$ & $-1.55$ \\
2026-11-04 14:57 & fixed & $+3.96$ & $+4.13$ & $+0.17$ \\
2026-11-04 15:00 & transpiler & $+0.96$ & $+1.74$ & $+0.78$ \\
\hline
\end{tabular}
\end{table}

The timescale of the drift is not fixed. In the first campaign agreement was
observed over up to $\hwVerifiedSpan$ minutes, and a run $\hwDriftSeenH$ hours
after the second session already differed; in the November campaign the
splitting moved within $\hwFixHourGapMin$ minutes. ``Within a session'' therefore
means ``within $\hwVerifiedSpan$ minutes in the first campaign''; it guarantees
no agreement in general. Table~\ref{tab:hw_gaps} lists the time gaps of all
comparisons; three are safe by construction (the mitigation ablation, the
alternating processor runs and the single-job control model). The
$\hwDevPct\,\%$ of Eq.~\eqref{eq:hw_dE} is the closest of the $\hwRepN$ runs to
the exact value; the claim of the paper rests on Eq.~\eqref{eq:hw_repro}.

One further run on a third day is excluded: its subspace population rises by
$\hwExclRiseSigma\,\sigma$ between consecutive depths, which leakage cannot do,
against at most $\hwRiseMaxOther\,\sigma$ in every other run. The criterion was formulated after this
run had been flagged, but it does not use the exact result. Including the run raises the spread from
$\hwRepRelPct\,\%$ to $\hwRepRelPctExcl\,\%$.

\begin{table}[htbp]
\centering
\caption{Elapsed time between the runs each comparison in Section~\ref{sec:hw_campaign} rests on. The threshold is not chosen: it is the 104~minutes over which repeated runs were \emph{observed} to agree to shot noise. Comparisons within it are marked \emph{verified}; the rest fall in the window between that span and the 7.4~hours after which drift was seen, where we have no measurement either way.}
\label{tab:hw_gaps}
\begin{tabular}{lrl}
\hline
Comparison & Gap [min] & Reproducibility \\
\hline
mitigation ablation vs its reference & 0 & verified \\
two processors, joint session & 13 & verified \\
signature scan vs its lambda=4 point & 94 & verified \\
post-selection validity, the two runs & 133 & unresolved \\
ZNE vs sampler & 166 & unresolved \\
control model vs unary retention & 292 & unresolved \\
\hline
\end{tabular}
\end{table}

Dynamical decoupling and readout twirling run in every measurement. Repeating
the protocol with both switched off, within a second of a repeat run, gives a
deviation in $\Delta E$ of $\hwAblDevPct\,\%$,
an amplitude of $\hwAblAmp$ and a surviving-signal scale of $n_0 = \hwAblNzero$
two-qubit gates, against $\hwAblDevPctOn\,\%$, $\hwAblAmpOn$ and
$\hwAblNzeroOn$ for its paired reference.

All three favour the unmitigated run, but the differences are of the size of
the run-to-run spread: the two mitigated runs, ten hours apart, differ from each
other by more. On this circuit the contribution of the mitigation is below the
reproducibility floor.

\section{Control Model on Hardware}\label{app:hw_control}

Subspace normalisation relies on the conserved excitation number: population
outside the one-excitation sector is noise and can be measured. To test whether
the gain belongs to the encoding or merely to a robust observable, we ran the
Ising chain of Appendix~\ref{sec:tier_f}, in which every bitstring is a legal
state, on the same device. Quenching from the polarised product state, the false
vacuum for $h>0$, we measure the magnetisation $m_z(t)$ on \hwIsBackend\ at $N=\hwIsN$, $J=\hwIsJ$, $g=\hwIsG$, $h=\hwIsH$ and
$\delta t=\hwIsDt$, out to $t=\hwIsTmax$ with $\hwIsShots$ shots per circuit,
over the same number of Trotter layers as the unary run of
Eq.~\eqref{eq:hw_dE}.
One layer costs $\hwIsTwoQPerStep$ two-qubit gates in both models; the Ising
circuits are $\hwIsTwoQOffset$ gates shallower throughout, since they need no
preparation cascade.

Noise drives $m_z$ towards zero, which is also where false-vacuum decay drives
it, so a measured decay establishes nothing by itself. The degenerate chain
$h=0$ stays near $m_z=\hwIsMzCtlIdeal$ instead and separates the two causes; it
was submitted in the same job as the $h=\hwIsH$ run, on the same qubits. At the
deepest point, $\hwIsTwoQMax$ two-qubit gates, the false vacuum reads
$m_z=\hwIsMzFv$ against a noiseless $\hwIsMzFvIdeal$, while the control reads
$\hwIsMzCtl$ against a noiseless $\hwIsMzCtlIdeal$. The two are
$\hwIsSepSigma\,\sigma$ apart on shot noise, so the decay is caused by the
longitudinal field, not the device. At that depth the decaying state occupies
$\hwIsIprFv$ bitstrings against $\hwIsIprCtl$ for the control, a factor
$\hwIsIprRatio$, as nucleated bubbles spread it through the computational basis.
The domain-wall density from the same shots reads $\hwIsDwFv$ (noiseless
$\hwIsDwFvIdeal$) for the false vacuum and $\hwIsDwCtl$ ($\hwIsDwCtlIdeal$) for
the control; noise pushes both up, but the ordering survives at every depth.

\begin{figure}[htbp]
\centering
\includegraphics[width=\textwidth]{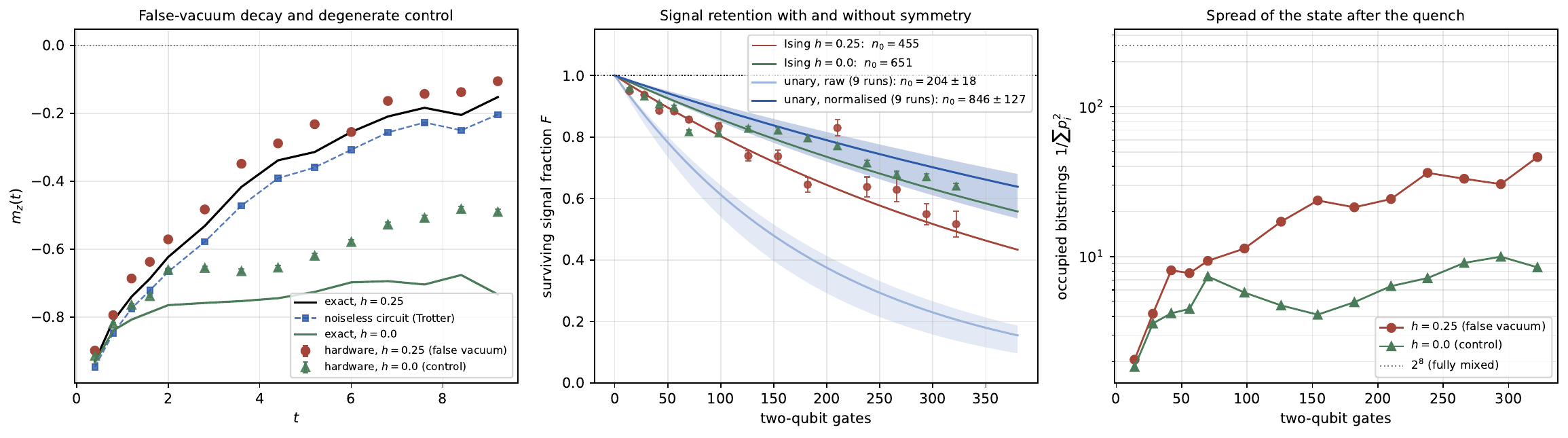}
\caption{\textbf{Control model on the same processor} (\hwIsBackend). \textbf{Left:} magnetisation after the quench.
Red, the false vacuum at $h=\hwIsH$; green, the degenerate chain $h=0$ at the
same depth in the same job. Lines: exact evolution; blue squares: noiseless circuit at $\delta t = \hwIsDt$.
\textbf{Centre:} surviving signal fraction $F$ (Eq.~\eqref{eq:hw_F}) against
two-qubit gate count. Ising runs: single measurements with fits. Unary
estimator with and without normalisation: bands over the $\hwRepN$ repetitions,
mean as line; their ratio, formed inside each run, reproduces to
$\hwIsRatioRelPct\,\%$. \textbf{Right:} number of occupied bitstrings, $1/\sum_i p_i^2$.}
\label{fig:hw_ising}
\end{figure}

Under a global
depolarising channel an observable contracts as $\langle O\rangle =
F\,\langle O\rangle_{\text{ideal}} + (1-F)\,\langle O\rangle_{\text{mixed}}$,
and the mixed-state value is not a free parameter: it is $0$ for $m_z$,
$2^{-N}$ for the probability of one bitstring, and $1/N$ for the normalised
ratio. $F$ can therefore be read off at every point without a fit, and its
decay with circuit size,
\begin{equation}
F(n_{2Q}) = \exp(-n_{2Q}/n_0),
\label{eq:hw_F}
\end{equation}
gives the gate count $n_0$ at which a share $1/e$ of the signal remains. The
two Ising runs give
$n_0 = \hwIsNzeroCtl \pm \hwIsNzeroCtlErr$ and
$\hwIsNzeroFv \pm \hwIsNzeroFvErr$ (Figure~\ref{fig:hw_ising}), against
$\hwIsNzeroPostMean \pm \hwIsNzeroPostSd$ for the normalised unary estimator over
the $\hwRepN$ repetitions (the band of the same figure) and
$\hwIsNzeroRawMean \pm \hwIsNzeroRawSd$ for the same data without
normalisation.

Only the gain from the normalisation is a safe comparison: it is a ratio of
two estimates from the same job and reproduces as
$\hwIsRatioMean \pm \hwIsRatioSd$ across the $\hwRepN$ repetitions (spread
$\hwIsRatioRelPct\,\%$). The absolute retention scatters by
$\hwIsNzeroPostRelPct\,\%$ and falls below the Ising control in $\hwIsNbelow$ of
the $\hwRepN$ runs, so the comparison between the models, taken hours or weeks apart, is
indicative only. The Ising model sits well above the unmitigated unary level
because $m_z$ averages $N$ single-qubit expectations, whereas $P_{\text{raw}}$
is the probability of one bitstring out of $2^N$. Such robustness can be bought
by choosing a bulk observable; the conserved quantity instead allows the leaked
population to be measured and divided out.

Without a symmetry, dividing the $h=\hwIsH$ curve point by point by the
contraction of the $h=0$ control also works: the root-mean-square deviation from
the noiseless circuit falls from $\hwIsRmsRaw$ to $\hwIsRmsCorr$, a factor
$\hwIsRmsFactor$. But it costs a second run and assumes, untestably, that the
noise acted identically on two different states. Circuit~B of
Section~\ref{sec:hw_campaign} instead measures the same state's own subspace
population, so the assumption can be checked within the measurement; here it is
tested only indirectly (Section~\ref{sec:hw_campaign}).

\noindent\textbf{Limits of the retention comparison.} The two Ising runs should return the
same $n_0$, since $F$ is meant to be a property of the circuit, not of the
state. Under the \texttt{FakeTorino} noise model they do, to within their
errors. On the device they do not: $\hwIsNzeroFv$ against $\hwIsNzeroCtl$, far
outside the quoted uncertainties. Two causes are consistent with the data: an additive bias from $T_1$ decay,
which enters the $h=\hwIsH$ ratio about four times more strongly (a free
asymptote brings the two $n_0$ within $1.5\,\sigma$), or a contraction that
depends on how far the state has spread (participation ratio larger by
$\hwIsIprRatio$). Either way a reference run is an empirical correction, not a
noise model, and Eq.~\eqref{eq:hw_F} summarises these curves rather than
characterising the device. The fit residual is $\hwIsResidCtl$ for the control
but $\hwIsResidUnRaw$ for the unmitigated unary estimator.

\end{appendices}

\bibliography{refs}

@article{garg_2000,
  author  = {Garg, Anupam},
  title   = {Tunnel splittings for one-dimensional potential wells revisited},
  journal = {American Journal of Physics},
  volume  = {68},
  number  = {5},
  pages   = {430--437},
  year    = {2000},
  doi     = {10.1119/1.19458}
}

@article{abel_2021,
  author  = {Abel, Steven and Chancellor, Nicholas and Spannowsky, Michael},
  title   = {Quantum computing for quantum tunneling},
  journal = {Physical Review D},
  volume  = {103},
  number  = {1},
  pages   = {016008},
  year    = {2021},
  doi     = {10.1103/PhysRevD.103.016008}
}

@article{cheng_2024,
  author  = {Cheng, Zhi-Jie and Wu, Yu-Kai and Li, Shi-Jiao and Mei, Quan-Xin and Li, Bo-Wen and Wang, Gang-Xi and Jiang, Yue and Qi, Bin-Xiang and Zhou, Zi-Chao and Hou, Pan-Yu and Duan, Lu-Ming},
  title   = {Observation of quantum superposition of topological defects in a trapped-ion quantum simulator},
  journal = {Science Advances},
  volume  = {10},
  number  = {42},
  pages   = {{eadr9527}},
  year    = {2024},
  doi     = {10.1126/sciadv.adr9527}
}

@article{lagnese_2021,
  author  = {Lagnese, Gianluca and Surace, Federica Maria and Kormos, M\'arton and Calabrese, Pasquale},
  title   = {False vacuum decay in quantum spin chains},
  journal = {Physical Review B},
  volume  = {104},
  number  = {20},
  pages   = {{L201106}},
  year    = {2021},
  doi     = {10.1103/PhysRevB.104.L201106}
}

@article{milsted_2022,
  author  = {Milsted, Ashley and Liu, Junyu and Preskill, John and Vidal, Guifre},
  title   = {Collisions of False-Vacuum Bubble Walls in a Quantum Spin Chain},
  journal = {PRX Quantum},
  volume  = {3},
  pages   = {020316},
  year    = {2022},
  doi     = {10.1103/PRXQuantum.3.020316}
}

@article{kormos_2017,
  author  = {Kormos, M\'arton and Collura, Mario and Tak\'acs, G\'abor and Calabrese, Pasquale},
  title   = {Real-time confinement following a quantum quench to a non-integrable model},
  journal = {Nature Physics},
  volume  = {13},
  pages   = {246--249},
  year    = {2017},
  doi     = {10.1038/nphys3934}
}

@article{jordan_2012,
  author  = {Jordan, Stephen P. and Lee, Keith S. M. and Preskill, John},
  title   = {Quantum Algorithms for Quantum Field Theories},
  journal = {Science},
  volume  = {336},
  number  = {6085},
  pages   = {1130--1133},
  year    = {2012},
  doi     = {10.1126/science.1217069}
}

@article{sornborger_2012,
  author  = {Sornborger, Andrew T.},
  title   = {Quantum Simulation of Tunneling in Small Systems},
  journal = {Scientific Reports},
  volume  = {2},
  pages   = {597},
  year    = {2012},
  doi     = {10.1038/srep00597}
}

@article{feng_2013,
  author  = {Feng, Guan-Ru and Lu, Yao and Hao, Liang and Zhang, Fei-Hao and Long, Gui-Lu},
  title   = {Experimental simulation of quantum tunneling in small systems},
  journal = {Scientific Reports},
  volume  = {3},
  pages   = {2232},
  year    = {2013},
  doi     = {10.1038/srep02232}
}

@misc{mcgarry_2026,
  author        = {McGarry, Cameron and Chalermpusitarak, Teerawat and Schwennicke, Kai and Scuccimarra, Frank and Millican, Maverick J. and Matsos, Vassili G. and Valahu, Christophe H. and Nagpal, Prachi and Chan, Hon-Kwan and Nourse, Henry L. and Kassal, Ivan and Tan, Ting Rei},
  title         = {Programmable quantum simulation of anharmonic dynamics},
  year          = {2026},
  eprint        = {2603.04744},
  archivePrefix = {arXiv},
  primaryClass  = {quant-ph},
  note          = {arXiv:2603.04744 [quant-ph]}
}

@article{escobar_ruiz_2015,
  author  = {Escobar-Ruiz, M. A. and Shuryak, E. and Turbiner, A. V.},
  title   = {Three-loop correction to the instanton density. {I}. {T}he quartic double well potential},
  journal = {Physical Review D},
  volume  = {92},
  pages   = {025046},
  year    = {2015},
  doi     = {10.1103/PhysRevD.92.025046}
}

@incollection{coleman1,
  author    = {Coleman, Sidney},
  title     = {The Uses of Instantons},
  booktitle = {The Whys of Subnuclear Physics},
  editor    = {Zichichi, Antonino},
  publisher = {Plenum Press},
  address   = {New York},
  pages     = {805--941},
  year      = {1979},
  doi       = {10.1007/978-1-4684-0991-8_16}
}

@article{vainshtein_abc,
  author  = {Vainshtein, A. I. and Zakharov, V. I. and Novikov, V. A. and Shifman, M. A.},
  title   = {{ABC} of instantons},
  journal = {Soviet Physics Uspekhi},
  volume  = {25},
  number  = {4},
  pages   = {195--215},
  year    = {1982},
  doi     = {10.1070/PU1982v025n04ABEH004533}
}

@book{zinn_justin_QFT,
  author    = {Zinn-Justin, Jean},
  title     = {Quantum Field Theory and Critical Phenomena},
  edition   = {4th},
  publisher = {Oxford University Press},
  address   = {Oxford},
  year      = {2002},
  doi       = {10.1093/acprof:oso/9780198509233.001.0001}
}

@article{schollwoeck_2011,
  author  = {Schollw\"ock, Ulrich},
  title   = {The density-matrix renormalization group in the age of matrix product states},
  journal = {Annals of Physics},
  volume  = {326},
  number  = {1},
  pages   = {96--192},
  year    = {2011},
  doi     = {10.1016/j.aop.2010.09.012}
}

@article{gray_quimb,
  author  = {Gray, Johnnie},
  title   = {quimb: A python package for quantum information and many-body calculations},
  journal = {Journal of Open Source Software},
  volume  = {3},
  number  = {29},
  pages   = {819},
  year    = {2018},
  doi     = {10.21105/joss.00819}
}

@article{raveh_nepomechie_2024,
  author  = {Raveh, David and Nepomechie, Rafael I.},
  title   = {Dicke states as matrix product states},
  journal = {Physical Review A},
  volume  = {110},
  pages   = {052438},
  year    = {2024},
  doi     = {10.1103/PhysRevA.110.052438}
}

@article{calabrese_cardy_2005,
  author  = {Calabrese, Pasquale and Cardy, John},
  title   = {Evolution of entanglement entropy in one-dimensional systems},
  journal = {Journal of Statistical Mechanics: Theory and Experiment},
  volume  = {2005},
  pages   = {P04010},
  year    = {2005},
  doi     = {10.1088/1742-5468/2005/04/P04010}
}

@article{bravyi_gosset_2016,
  author  = {Bravyi, Sergey and Gosset, David},
  title   = {Improved Classical Simulation of Quantum Circuits Dominated by {C}lifford Gates},
  journal = {Physical Review Letters},
  volume  = {116},
  pages   = {250501},
  year    = {2016},
  doi     = {10.1103/PhysRevLett.116.250501}
}

@article{leone_2022,
  author  = {Leone, Lorenzo and Oliviero, Salvatore F. E. and Hamma, Alioscia},
  title   = {Stabilizer {R}\'enyi Entropy},
  journal = {Physical Review Letters},
  volume  = {128},
  pages   = {050402},
  year    = {2022},
  doi     = {10.1103/PhysRevLett.128.050402}
}

@article{haug_piroli_2023,
  author  = {Haug, Tobias and Piroli, Lorenzo},
  title   = {Quantifying nonstabilizerness of matrix product states},
  journal = {Physical Review B},
  volume  = {107},
  pages   = {035148},
  year    = {2023},
  doi     = {10.1103/PhysRevB.107.035148}
}

@article{low_chuang_2019,
  author  = {Low, Guang Hao and Chuang, Isaac L.},
  title   = {Hamiltonian Simulation by Qubitization},
  journal = {Quantum},
  volume  = {3},
  pages   = {163},
  year    = {2019},
  doi     = {10.22331/q-2019-07-12-163}
}

@article{berry_2015,
  author  = {Berry, Dominic W. and Childs, Andrew M. and Cleve, Richard and Kothari, Robin and Somma, Rolando D.},
  title   = {Simulating {H}amiltonian Dynamics with a Truncated {T}aylor Series},
  journal = {Physical Review Letters},
  volume  = {114},
  pages   = {090502},
  year    = {2015},
  doi     = {10.1103/PhysRevLett.114.090502}
}

@article{childs_2021,
  author  = {Childs, Andrew M. and Su, Yuan and Tran, Minh C. and Wiebe, Nathan and Zhu, Shuchen},
  title   = {Theory of {T}rotter Error with Commutator Scaling},
  journal = {Physical Review X},
  volume  = {11},
  pages   = {011020},
  year    = {2021},
  doi     = {10.1103/PhysRevX.11.011020}
}

@article{rutkevich_1999,
  author  = {Rutkevich, S. B.},
  title   = {Decay of the metastable phase in $d=1$ and $d=2$ {I}sing models},
  journal = {Physical Review B},
  volume  = {60},
  pages   = {14525--14528},
  year    = {1999},
  doi     = {10.1103/PhysRevB.60.14525}
}

@article{rychkov_vitale_2016,
  author  = {Rychkov, Slava and Vitale, Lorenzo G.},
  title   = {Hamiltonian truncation study of the $\Phi^4$ theory in two dimensions. {II}. {T}he $\mathbb{Z}_2$-broken phase and the {C}hang duality},
  journal = {Physical Review D},
  volume  = {93},
  pages   = {065014},
  year    = {2016},
  doi     = {10.1103/PhysRevD.93.065014}
}

@book{coleman_aspects,
  author    = {Coleman, Sidney},
  title     = {Aspects of Symmetry: Selected {E}rice Lectures},
  publisher = {Cambridge University Press},
  address   = {Cambridge},
  year      = {1985},
  doi       = {10.1017/CBO9780511565045}
}

@article{wohler_shuryak_1994,
  author  = {W\"ohler, C. F. and Shuryak, E. V.},
  title   = {Two-loop correction to the instanton density for the double well potential},
  journal = {Physics Letters B},
  volume  = {333},
  pages   = {467--470},
  year    = {1994},
  doi     = {10.1016/0370-2693(94)90169-4}
}

@article{bonet_monroig_2018,
  author  = {Bonet-Monroig, X. and Sagastizabal, R. and Singh, M. and O'Brien, T. E.},
  title   = {Low-cost error mitigation by symmetry verification},
  journal = {Physical Review A},
  volume  = {98},
  pages   = {062339},
  year    = {2018},
  doi     = {10.1103/PhysRevA.98.062339}
}

@article{mcardle_2019,
  author  = {McArdle, Sam and Yuan, Xiao and Benjamin, Simon},
  title   = {Error-Mitigated Digital Quantum Simulation},
  journal = {Physical Review Letters},
  volume  = {122},
  pages   = {180501},
  year    = {2019},
  doi     = {10.1103/PhysRevLett.122.180501}
}

@article{sagastizabal_2019,
  author  = {Sagastizabal, R. and Bonet-Monroig, X. and Singh, M. and Rol, M. A. and Bultink, C. C. and Fu, X. and Price, C. H. and Ostroukh, V. P. and Muthusubramanian, N. and Bruno, A. and Beekman, M. and Haider, N. and O'Brien, T. E. and DiCarlo, L.},
  title   = {Experimental error mitigation via symmetry verification in a variational quantum eigensolver},
  journal = {Physical Review A},
  volume  = {100},
  pages   = {010302},
  year    = {2019},
  doi     = {10.1103/PhysRevA.100.010302}
}

@article{sawaya_2020,
  author  = {Sawaya, Nicolas P. D. and Menke, Tim and Kyaw, Thi Ha and Johri, Sonika and Aspuru-Guzik, Al\'an and Guerreschi, Gian Giacomo},
  title   = {Resource-efficient digital quantum simulation of d-level systems for photonic, vibrational, and spin-s {H}amiltonians},
  journal = {npj Quantum Information},
  volume  = {6},
  pages   = {49},
  year    = {2020},
  doi     = {10.1038/s41534-020-0278-0}
}

@article{di_matteo_2021,
  author  = {Di Matteo, Olivia and McCoy, Anna and Gysbers, Peter and Miyagi, Takayuki and Woloshyn, R. M. and Navr\'atil, Petr},
  title   = {Improving {H}amiltonian encodings with the {G}ray code},
  journal = {Physical Review A},
  volume  = {103},
  pages   = {042405},
  year    = {2021},
  doi     = {10.1103/PhysRevA.103.042405}
}

@article{ramos_calderer_2021,
  author  = {Ramos-Calderer, Sergi and P\'erez-Salinas, Adri\'an and Garc\'ia-Mart\'in, Diego and Bravo-Prieto, Carlos and Cortada, Jorge and Planagum\`a, Jordi and Latorre, Jos\'e I.},
  title   = {Quantum unary approach to option pricing},
  journal = {Physical Review A},
  volume  = {103},
  pages   = {032414},
  year    = {2021},
  doi     = {10.1103/PhysRevA.103.032414}
}

@article{kivlichan_2018,
  author  = {Kivlichan, Ian D. and McClean, Jarrod and Wiebe, Nathan and Gidney, Craig and Aspuru-Guzik, Al\'an and Chan, Garnet Kin-Lic and Babbush, Ryan},
  title   = {Quantum Simulation of Electronic Structure with Linear Depth and Connectivity},
  journal = {Physical Review Letters},
  volume  = {120},
  pages   = {110501},
  year    = {2018},
  doi     = {10.1103/PhysRevLett.120.110501}
}

@article{arrazola_2022,
  author  = {Arrazola, Juan Miguel and Di Matteo, Olivia and Quesada, Nicol\'as and Jahangiri, Soran and Delgado, Alain and Killoran, Nathan},
  title   = {Universal quantum circuits for quantum chemistry},
  journal = {Quantum},
  volume  = {6},
  pages   = {742},
  year    = {2022},
  doi     = {10.22331/q-2022-06-20-742}
}

@article{jahnke_lubich_2000,
  author  = {Jahnke, Tobias and Lubich, Christian},
  title   = {Error Bounds for Exponential Operator Splittings},
  journal = {BIT Numerical Mathematics},
  volume  = {40},
  number  = {4},
  pages   = {735--744},
  year    = {2000},
  doi     = {10.1023/A:1022396519656}
}

@article{childs_realspace_2022,
  author  = {Childs, Andrew M. and Leng, Jiaqi and Li, Tongyang and Liu, Jin-Peng and Zhang, Chenyi},
  title   = {Quantum simulation of real-space dynamics},
  journal = {Quantum},
  volume  = {6},
  pages   = {860},
  year    = {2022},
  doi     = {10.22331/q-2022-11-17-860}
}

@article{zalka_1998,
  author  = {Zalka, Christof},
  title   = {Simulating quantum systems on a quantum computer},
  journal = {Proceedings of the Royal Society of London. Series A},
  volume  = {454},
  number  = {1969},
  pages   = {313--322},
  year    = {1998},
  doi     = {10.1098/rspa.1998.0162}
}

@article{kassal_2008,
  author  = {Kassal, Ivan and Jordan, Stephen P. and Love, Peter J. and Mohseni, Masoud and Aspuru-Guzik, Al\'an},
  title   = {Polynomial-time quantum algorithm for the simulation of chemical dynamics},
  journal = {Proceedings of the National Academy of Sciences},
  volume  = {105},
  number  = {48},
  pages   = {18681--18686},
  year    = {2008},
  doi     = {10.1073/pnas.0808245105}
}

@article{benenti_strini_2008,
  author  = {Benenti, Giuliano and Strini, Giuliano},
  title   = {Quantum simulation of the single-particle {S}chr\"odinger equation},
  journal = {American Journal of Physics},
  volume  = {76},
  number  = {7},
  pages   = {657--662},
  year    = {2008},
  doi     = {10.1119/1.2894532}
}

@article{macridin_2018,
  author  = {Macridin, Alexandru and Spentzouris, Panagiotis and Amundson, James and Harnik, Roni},
  title   = {Digital quantum computation of fermion-boson interacting systems},
  journal = {Physical Review A},
  volume  = {98},
  number  = {4},
  pages   = {042312},
  year    = {2018},
  doi     = {10.1103/PhysRevA.98.042312}
}

@article{klco_savage_2019,
  author  = {Klco, Natalie and Savage, Martin J.},
  title   = {Digitization of scalar fields for quantum computing},
  journal = {Physical Review A},
  volume  = {99},
  number  = {5},
  pages   = {052335},
  year    = {2019},
  doi     = {10.1103/PhysRevA.99.052335}
}

@article{clements_2016,
  author  = {Clements, William R. and Humphreys, Peter C. and Metcalf, Benjamin J. and Kolthammer, W. Steven and Walmsley, Ian A.},
  title   = {Optimal design for universal multiport interferometers},
  journal = {Optica},
  volume  = {3},
  number  = {12},
  pages   = {1460},
  year    = {2016},
  doi     = {10.1364/OPTICA.3.001460}
}

@article{an_fang_lin_2021,
  author  = {An, Dong and Fang, Di and Lin, Lin},
  title   = {Time-dependent unbounded {H}amiltonian simulation with vector norm scaling},
  journal = {Quantum},
  volume  = {5},
  pages   = {459},
  year    = {2021},
  doi     = {10.22331/q-2021-05-26-459}
}

@article{childs_su_2019,
  author  = {Childs, Andrew M. and Su, Yuan},
  title   = {Nearly Optimal Lattice Simulation by Product Formulas},
  journal = {Physical Review Letters},
  volume  = {123},
  pages   = {050503},
  year    = {2019},
  doi     = {10.1103/PhysRevLett.123.050503}
}

@article{burgarth_2024,
  author  = {Burgarth, Daniel and Facchi, Paolo and Hahn, Alexander and Johnsson, Mattias and Yuasa, Kazuya},
  title   = {Strong error bounds for {T}rotter and {S}trang-splittings and their implications for quantum chemistry},
  journal = {Physical Review Research},
  volume  = {6},
  pages   = {043155},
  year    = {2024},
  doi     = {10.1103/PhysRevResearch.6.043155}
}

@misc{wiesner_1996,
  author        = {Wiesner, Stephen},
  title         = {Simulations of Many-Body Quantum Systems by a Quantum Computer},
  year          = {1996},
  eprint        = {quant-ph/9603028},
  archivePrefix = {arXiv},
  primaryClass  = {quant-ph},
  note          = {arXiv:quant-ph/9603028}
}

@article{vodeb_2025,
  author  = {Vodeb, Jaka and Desaules, Jean-Yves and Hallam, Andrew and Rava, Andrea and Humar, Gregor and Willsch, Dennis and Jin, Fengping and Willsch, Madita and Michielsen, Kristel and Papi\'c, Zlatko},
  title   = {Stirring the false vacuum via interacting quantized bubbles on a 5,564-qubit quantum annealer},
  journal = {Nature Physics},
  volume  = {21},
  number  = {3},
  pages   = {386--392},
  year    = {2025},
  doi     = {10.1038/s41567-024-02765-w}
}

@article{zhu_2026,
  author  = {Zhu, Zi-Hang and Liu, Ying and Lagnese, Gianluca and Surace, Federica Maria and Zhang, Wei-Yong and He, Ming-Gen and Halimeh, Jad C. and Dalmonte, Marcello and Morampudi, Siddhardh C. and Wilczek, Frank and Yuan, Zhen-Sheng and Pan, Jian-Wei},
  title   = {Probing False Vacuum Decay on a Cold-Atom Gauge-Theory Quantum Simulator},
  journal = {Physical Review Letters},
  volume  = {137},
  number  = {14},
  pages   = {141601},
  year    = {2026},
  doi     = {10.1103/vhnm-j9bw}
}

\end{document}